\documentclass[pra,preprint,11pt,superscriptaddress]{revtex4-1} 
\usepackage{amsmath}
\usepackage{amssymb}
\usepackage{graphicx}
\usepackage{color}
\newcommand{\be}[1]{\begin{equation}\label{#1}}
\newcommand{\ee}{\end{equation}}
\newcommand{\bea}[1]{\begin{eqnarray}\label{#1}}
\newcommand{\eea}{\end{eqnarray}}
\newcommand{\no}{\nonumber \\}
\newcommand{\Fig}[1]{Fig.(\ref{#1})}

\newcommand{\Eq}[1]{Eq.(\ref{#1})}

\newcommand{\Sec}[1]{Section~\ref{#1}}
\newcommand{\bsub}{\begin{subequations}}
\newcommand{\esub}{\end{subequations}}

\newcommand{\fracd}[2]{\frac{\displaystyle #1}{\displaystyle #2}}
\newcommand{\om}{\omega}

\def\a0{{\alpha_0}}

\def\da0{{\dot{\alpha}_0}}

\def\myover#1{\myoverDefn#1}
\def\myoverDefn#1#2{\hbox{\space \raise-2mm\hbox{$\textstyle{#1} \atop \scriptstyle{#2}$} }}

\def\om{{\omega}}

\def\mA{{\mathcal{A}}}
\def\mB{{\mathcal{B}}}
\def\mC{{\mathcal{C}}}

\def\g{{\gamma}}

\def\a{{\alpha}}

\def\dag{\dagger}

\def\exia{e^{i\xi_a L}}
\def\exib{e^{i\xi_b L}}
\def\exik{e^{i\xi_k L}}

\def\rp{r_{P}}
\def\rp2{r_{p}^{2}}
\def\Tr{\textrm{Tr}}

\def\tH{\tilde{H}}
\def\tf{\tilde{f}}
\def\thaa{\theta_{G_{aa}}}
\def\thbb{\theta_{G_{bb}}}
\def\tA{\tilde{A}}
\def\tB{\tilde{B}}
\def\tX{\tilde{X}}
\def\tnu{\tilde{\nu}}
\def\kinab{k\in\{a,b\}}
\newcommand{\half}{\frac{1}{2}}

\newcommand{\ket}[1]{|#1\rangle}
\newcommand{\bra}[1]{\langle #1|}

\begin{document}
\title{Photon pair generation in a lossy microring resonator. II. Entanglement in the output mixed Gaussian squeezed state}
\author{Paul M. Alsing}
\affiliation{Air Force Research Laboratory, Information Directorate, 525 Brooks Rd, Rome, NY, 13411}
\author{Edwin E. Hach III}
\affiliation{Rochester Institute of Technology, School of Physics and Astronomy, 85 Lomb Memorial Dr., Rochester, NY 14623}
\date{\today}

\begin{abstract}
%
In this work we examine the entanglement of the output signal-idler squeezed vacuum state in the Heisenberg picture as a function
of the coupling and internal propagation loss parameters of a microring resonator.
Using the log-negativity as a measure of entanglement for a mixed Gaussian state,
we examine the competitive effects of the transfer matrix that encodes the classical phenomenological loss, as well as the matrix that
that incorporates the coupling and internal propagation loss due to the quantum Langevin  noise fields  required to preserve unitarity of the composite system,
(signal-idler) and environment (noise) structure.
\end{abstract}
\maketitle
\section{Introduction}\label{sec:intro}
In the first paper of this two-part investigation ``Photon pair generation in a lossy microring resonator. I. Theory," \cite{Alsing_Hach:2017a} (designated AH-I) we developed the theory for entangled photon pair generation in a microring resonator (mrr) using a recent input-output formalism based on the work of
Raymer and McKinstrie  \cite{Raymer:2013} and Alsing, \textit{et al.} \cite{Alsing_Hach:2016}
that incorporates the circulation factors that account for the multiple round trips  of the fields within the cavity.
In AH-I we considered biphoton pair generation within the mrr via both SPDC and SFWM processes, and
computed the generated two-photon signal-idler intra-cavity and output state from a single bus (all-through) mrr.
In addition, we also computed the two-photon generation, coincidence-to-accidental, heralding efficiency rates,
and compared our results to related derivations of the Schrodinger picture biphoton state \cite{Scholz:2009, Tsang:2011, Sipe:2015a}  obtained using the
standard Langevin input-output formalism.

In this work, we examine entanglement of the output signal-idler squeezed vacuum state from the mrr in the Heisenberg picture as a function
of its coupling and internal propagation loss parameters. The squeezed output fields arising from either SPDC or SFWM generated within the mrr contain two types of terms:
(i) a transfer matrix that encodes the classical phenomenological loss and
(ii) a matrix that incorporates the coupling and internal propagation loss due to the quantum Langevin  noise fields
that are required to preserve unitarity of the composite system, (signal-idler) and environment (noise) structure.
Using the log-negativity as a measure of entanglement for a lossy Gaussian state,
we examine the competitive effects both of these terms as a function of the mrr loss parameters.
Authors such as Agarwal \textit{et al.} \cite{Agarwal:2010a,Agarwal:2013} have investigated the entanglement of two-mode mixed Gaussian states using the log negativity, while authors such as Sipe \textit{et al.} \cite{Sipe:2015a,Sipe:2015b} have investigated loss in a mrr. However, to our knowledge, the work presented here represents the first investigation of the entanglement of the squeezed output of a lossy mrr as a function of the parameters of this passive feedback device.

This paper is organized as follows.
%
In \Sec{sec:SPDC:SFWM:processes} we briefly review the main results of AH-I for the output operators for a lossy single-bus mrr given the input driving fields.
%
In \Sec{sec:squeezed:state} we examine the form of the output squeezed vacuum state from a Heisenberg operator perspective. This allows us to form the operator that generates the output squeezed  state, as well as the unitary operator that evolves the external input operators to the output operators, in the presence of loss. We further examine the entanglement of the output squeezed state employing the log negativity and explore its dependence on the mrr coupling and internal propagation loss.
%
In \Sec{sec:summary:discussion} we summarize our results and indicate avenues for future research.

\section{A brief review of AH-I: SPDC and SFWM processes inside a (single bus) microring resonator}\label{sec:SPDC:SFWM:processes}
\subsection{Preliminaries}\label{subsec:SPDC:SFWM:preliminaries}
\begin{figure}[h]
\includegraphics[width=3.0in,height=2.5in]{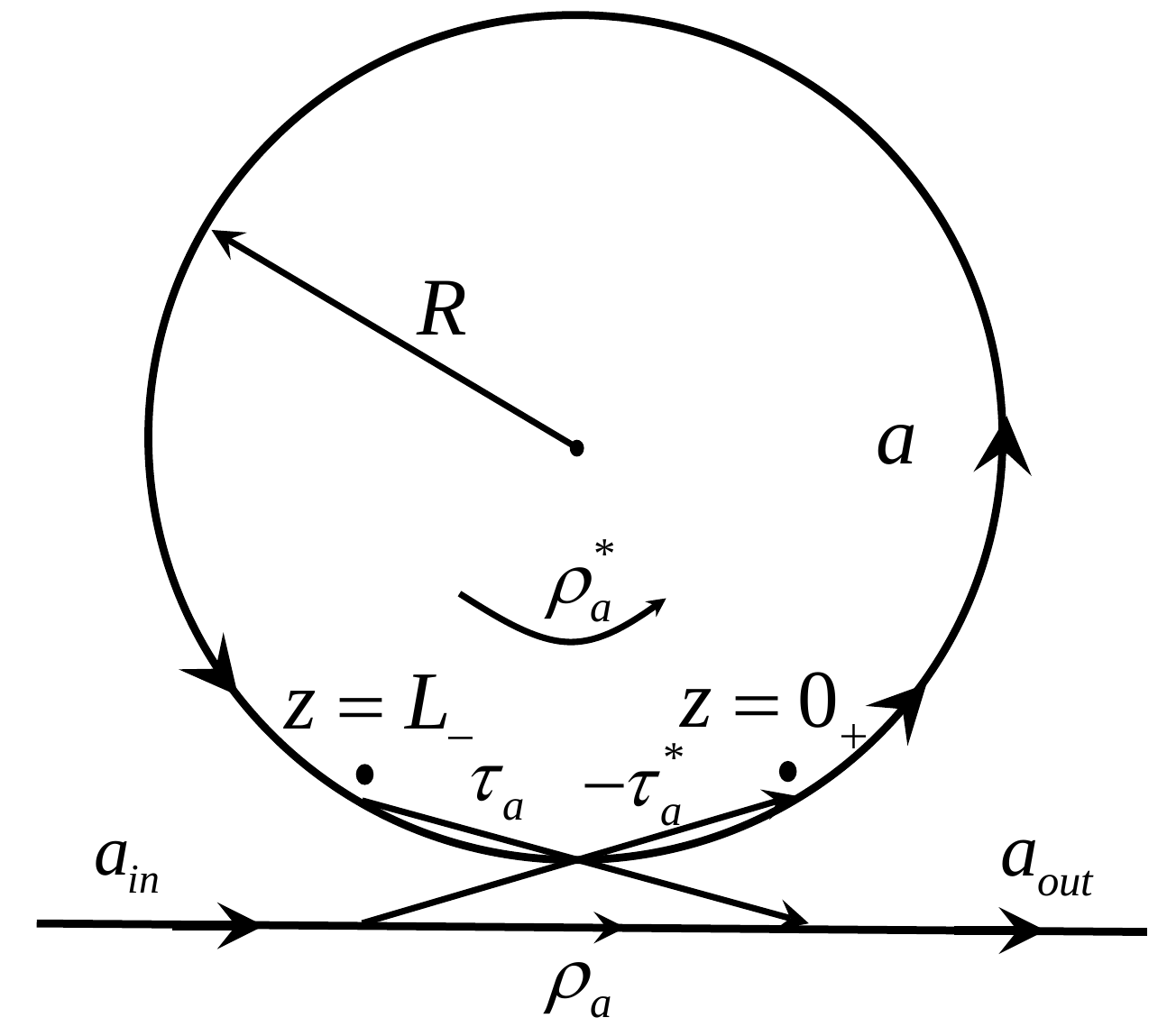}
\caption{A single bus (all-through) microring resonator (mrr) of length $L=2\pi R$ with intracavity field $a$, coupled to a waveguide bus with input field $a_{in}$ and output field $a_{out}$. $\rho_a\, \tau_a$ are the beam splitter like self-coupling and cross-coupling strengths, respectively, of the bus to the mrr such that $|\rho_a|^2 + |\tau_a|^2 =1$. $z=0_+$ is the point just inside the mrr which cross-couples to the input field $a_{in}$, and $z=L_-$ is the point after one round trip in the mrr that cross-couples to the output field $a_{out}$.
}\label{fig:single_bus_rr}
\end{figure}

Consider a single bus microring resonator (mrr) of length $L=2\pi R$, as illustrated in \Fig{fig:single_bus_rr}.
Here, $a$ is the intracavity field which is coupled to a waveguide bus with input field $a_{in}$ and output field $a_{out}$.
The parameters $\rho_a,\,\tau_a$ are the beam splitter like self-coupling and cross-coupling strengths, respectively, of the bus to the mrr such that $|\rho_a|^2 + |\tau_a|^2 =1$. $z=0_+$ is the point just inside the mrr which cross-couples to the input field $a_{in}$, and $z=L_-$ is the point after on round trip in the mrr that cross-couples to the output field $a_{out}$.

In the work of Raymer and McKinstrie \cite{Raymer:2013}  an internal cavity field $a$ satisfies a
traveling-wave Maxwell ODE in the absence of internal propagation loss  given by
\be{RM:1:v3}
(\partial_t + v_a\,\partial_z)\, a(z,t) = \alpha_{polz}\,P(z,t),
\ee
where $a(z,t)$ is the ring resonator cavity field (in the time domain), $v_a$ is the group velocity, $P(z,t)$ is the polarization and $\alpha_{polz}$ is a coupling constant. The carrier wave frequency has been factored out so that all frequencies are relative to the optical carrier frequency. The input coupling and periodicity of the cavity are captured by the boundary conditions
\bsub
\bea{RM:2:v3}
 a(0_+,t) &=& \rho_a\,a(L_-,t) + \tau_a\, a_{in}(t), \\
 a_{out}(t)  &=& \tau_a\,a(L_-,t) - \rho_a\, a_{in}(t), \label{RM:3:v3}
\eea
\esub
where we have taken the  beam splitter like  self-coupling $\rho_a$ (buss-bus, mrr-mrr),
and cross-coupling $\tau_a$ (bus-mrr) real for simplicity
and the minus sign in \Eq{RM:3:v3} accounts for the $\pi$ change in phase arising from the "reflection" of the input field off the higher index of refraction mrr to the output (bus) field.
The input and output fields satisfy the free field commutators
 \be{RM:5:34}
 [ a_{in}(t), a_{in}^\dagger(t')] = \delta(t-t') = [ a_{out}(t), a_{out}^\dagger(t')].
\ee

In the absence of loss, the above equations in the Fourier domain yield the unimodular transfer function $G_{out, in}(\om)$ defined by \cite{Raymer:2013}
\be{RM:16:17:v3}
 a_{out}(\om) \equiv  G_{out, in}(\om) \, a_{in}(\om), \qquad G_{out, in}(\om) =  e^{i\om T_a}\,\left[\frac{1-\rho_a\,e^{-i\om T_a}}{1-\rho_a\,e^{i\om T_a}}\right],
 \qquad |G_{out, in}(\om)| = 1.
\ee
With the inclusion of internal propagation loss, Alsing \textit{et al.} \cite{Alsing_Hach:2016} obtain the form
\bsub
\bea{aout:AH:v3}
a_{out}(\om) &=& G_{out,in}(\om)\,a_{in} + H_{out,in}(\om)\,f_a(\om), \\
G_{out,in}(\om) &=&
\left(
\frac{\rho_a-\alpha_a\,e^{i\theta_a}}{1-\rho^*_a\,\alpha_a\,e^{i\theta_a}}
\right),
\qquad |H_{out,in}(\om)| = \sqrt{1 - |G_{out,in}(\om)|^2}, \label{aout:AH:v3:G_and_H}
\eea
\esub
which defines the quantum noise operator $f_a(\om)$ from the requirement of the preservation of the free field output commutator.
Here, $G_{out,in}(\om)$ has the same form as the semi-classical case (see e.g. Yariv\cite{Yariv:2000}, and Rabus \cite{Rabus:2007})
that one obtains with the inclusion of a phenomenological loss factor $0\le\alpha_a\le1$ (see also \cite{Hach:2014}).

\subsection{Biphoton generation within the mrr}\label{subsec:preliminaries_biphoton:generation}
For biphoton generation arising from either a $\chi^{(2)}$ process of spontaneous parametric down conversion (SPDC), or a $\chi^{(3)}$ process of spontaneous four-wave mixing (SFWM), AH-I consider a signal mode $a$, and idler mode $b$ circulating within the mrr.
In the non-depleted pump approximation the authors consider the Hamiltonian
\bsub
\bea{H:NL}
 \mathcal{H}^{NL} &=&  \int_{-\infty}^{\infty}\,\frac{d\om}{2\pi}\,g(\om)\,
 \left( \alpha_p(z,\om)\, a^\dag(z,\om)\,b^\dag(z,\om) + \alpha^*_p(z,\om)\,a(z,\om)\,b(z,\om) \right), \\
g(\om) &=& g_{spcd}(\om)=\frac{3\,(\hbar\,\om_c)^{3/2}\,\chi^{(2)}}{4\epsilon_0\,\bar{n}^4\,V_{ring}},
 \quad \textrm{for SPDC}, \\
g(\om) &=& g_{sfwm}(\om) = \frac{3(\hbar\,\om_c)^2\,\chi^{(3)}}{4\epsilon_0\,\bar{n}^4\,V_{ring}},
\quad \textrm{for SFWM},
\eea
\esub
where  $\alpha_p$ is the complex c-number (constant) amplitude for the pump.
Thus, for each  of the nonlinear processes the signal and idler modes
satisfy the equation of motion  in the frequency domain
\bsub
\bea{a:b:EoM}
(-i\,\om + v_a\,\partial_z)\, a(z,\om) &=&  -i\,g\,\alpha_p(z,\om)\,b^\dag(z,\om) -\frac{\g'_a}{2}\,a(z,\om) + \alpha_{polz}\,f_a(z,\om), \label{a:EoM}\\
(-i\,\om + v_b\,\partial_z)\, b^\dag(z,\om) &=&   i\,g\,\alpha^*_p(z,\om)\,a(z,\om) -\frac{\g'_b}{2}\,b^\dag(z,\om) + \alpha_{polz}\,f_b(z,\om). \label{b:EoM}
\eea
\esub
Here $\g'_k$  is the internal propagation loss for mode $\kinab$, and $f_k$ are corresponding Langevin noise operators added to preserve the canonical form of the output commutators. Each mode $k$ satisfies its own pair of mrr input-output boundary conditions of the form of \Eq{RM:2:v3} and \Eq{RM:3:v3}.
By using these boundary conditions to eliminate the internal signal and idler cavity modes, and by defining
\be{defns:vectors}
 \vec{a}_{in}(\om) =
 \left(
   \begin{array}{c}
     a_{in}(\om) \\
     b_{in}^\dag(\om) \\
   \end{array}
 \right), \quad
 \vec{a}_{out}(\om) =
 \left(
   \begin{array}{c}
     a_{out}(\om) \\
     b_{out}^\dag(\om) \\
   \end{array}
 \right), \quad
\vec{f}(\om) =
 \left(
   \begin{array}{c}
     f_a(\om) \\
     f_b^\dag(\om) \\
   \end{array}
 \right). \qquad\qquad
\ee
AH-I obtained the following expression for the output fields in terms of the input fields and quantum noise operators
\be{aout:matrix:form}
\vec{a}_{out}(\om) = G(\om) \, \vec{a}_{in}(\om) + H(\om)\, \vec{f}(\om).
\ee
The expression for the matrices $G(\om)$ and $H(\om)$ are given by
\bsub
\bea{Gom:v2}
 G(\om) &=&
\left(
  \begin{array}{cc}
    G_{aa}(\om) & G_{ab}(\om) \\
    G_{ba}(\om) & G_{bb}(\om) \\
  \end{array}
\right), \label{Gom:v2:matrixform}\\
&\equiv& \frac{1}{D}
\left(
  \begin{array}{cc}
    (\exia - \rho_a )\,(1 - \rho_b\,\exib) + r_a\,r_b\,\rho_a  &   -i\,r_a\,\tau_a\,\tau_b\,\exib \\
    i\,r_b\,\tau_b\,\tau_a\,\exia       &  (\exib - \rho_b)\,(1 - \rho_a\,\exia) + r_a\,r_b\,\rho_b \\
  \end{array}
\right), \qquad \label{Gom:v2:matrix:elements}
\eea
\esub
with
\bsub
\bea{defns:S_D_dab:v2}
D &=& (1 - \rho_a\,\exia)\,(1 - \rho_b\,\exib) - r_a\,r_b, \qquad r_a = g\alpha_P T_a\, \quad \,r_b =g\alpha^*_P T_b,\\
\alpha_k &=& e^{-\gamma'_k/2\,T_k}, \quad \theta_k = \om\,T_k, \quad \textrm{for}\quad k\in\{a,b\},
\eea
\esub
and
\be{Hom:v2}
H(\om) =
\left(
  \begin{array}{cc}
    H_{aa}(\om) & H_{ab}(\om) \\
    H_{ba}(\om) & H_{bb}(\om) \\
  \end{array}
\right)
= \frac{1}{D} \,
\left(
  \begin{array}{cc}
    \tau_a\,(1-\rho_b\exib)  & -i\,r_a\,\tau_a \\
     i\,r_b\,\tau_b          & \tau_b\,(1-\rho_a\exia)
  \end{array}
\right). \qquad
\ee
In the above, we have defined $\exik \equiv \alpha_k \, e^{i\theta_k}$ with $\alpha_k = e^{-\gamma'_k\,T_k/2}$,
$\theta_k = \om\,T_k$, $r_a = g\,\alpha_p\,T_a$, and $r_b = g\,\alpha^*_p\,T_b$,
where $T_k=L/v_k$ is the cavity round trip time for mode $k\in\{a,b\}$.
To lowest order in the coupling $|g\alpha_p|$, we have $1/D\approx S_a\,S_b$
Note that to lowest order in $|g\alpha_p|$, we have $1/D\approx S_a\,S_b$
where
$S_k = \frac{1}{1-\rho_k\,\exik} = \sum_{n=0}^{\infty}\, \left(\rho_k\,\exik\right)^n \equiv \sum_{n=0}^{\infty}\, \left(\rho_k\,\alpha_k\,e^{i\,\theta_k}\right)^n$
for $k\in\{a,b\}$
are the geometric series  factors
resulting from the round trip circulations of the internal fields $k\in\{a,b\}$ inside the ring resonator.
Note that terms such as $(1 - \rho_a\,\exia)\,S_k$ are of the single-mode Yariv-form $G_{out,in}(\om)$ of \Eq{aout:AH:v3:G_and_H},
so that the diagonal terms $G_{k,k}$ in  \Eq{Gom:v2} represent their interacting, multi-mode generalization.
For typical ring resonator of radius $R=20 \mu$m and pump laser power of 1mW ($\chi^{(2)}\sim 2\times 10^{-12}$ m/V, $\alpha_p\sim 10^3$ V/m,) and round trip times of $T_k\sim$ 1 ps, we have $r_p\sim 10^{-5}$ \cite{Schneeloch:2017}.

\subsection{Commutators of the noise operators}\label{subsec:comm:noise:operators}
The commutation relations between the noise operators are fundamentally determined by the canonical commutators of the free input and output fields.
Given that the input fields satisfy $[a_{in}(\om), a^\dag_{in}(\om')] = [b_{in}(\om), b^\dag_{in}(\om')] = \delta(\om-\om')$, and that they each commute with the
noise operators $f_{a}(\om),\,f_{b}(\om)$ (via causality), one must also have that
$[a_{out}(\om), a^\dag_{out}(\om')] = [b_{out}(\om), b^\dag_{out}(\om')] = \delta(\om-\om')$.
Thus,  the requirement of independence and unitarity of the output field modes determines the set of linear equations
\bsub
\bea{comms:spdc}
{} [a_{out}(\om), a^\dag_{out}(\om')]&=& \delta(\om-\om') \Rightarrow  |H_{aa}|^2\, C_{aa} - |H_{ab}|^2\,C_{bb} + 2 \textrm{Re}( H_{aa}\,H^*_{ab}\,D_{ab})
= 1 - (|G_{aa}|^2 - |G_{ab}|^2), \qquad \label{comms:spdc:aadag}\\
{} [b_{out}(\om), b^\dag_{out}(\om')] &=& \delta(\om-\om') \Rightarrow  -|H_{ba}|^2\, C_{aa} + |H_{bb}|^2\,C_{bb} + 2 \textrm{Re}( H_{ba}\,H^*_{bb}\,D_{ab})
= 1 - (|G_{bb}|^2 - |G_{ba}|^2), \label{comms:spdc:bbdag} \qquad\quad\\
{} [a_{out}(\om), b_{out}(\om')] &=& 0 \Rightarrow  H_{aa} \, H^*_{ba} \, C_{aa} - H_{ab}\,H^*_{bb}\,C_{bb} +  H_{aa}\,H^*_{bb}\,D_{ab} + H_{ab}\,H^*_{ba}\,D^*_{ab}
= G_{ab}\,G^*_{bb} - G_{aa}\,G^*_{ba}, \qquad \label{comms:spdc:ab}\\
{} [a_{out}(\om), b^\dag_{out}(\om')] &=& 0 \Rightarrow \textrm{det}(H) \label{comms:spdc:abdag}\, C_{ab} = 0,
\eea
\esub
for the four constants $C_{aa}$, $C_{bb}$, $C_{ab}$, $D_{ab}$ defined by the commutation relations
\bsub
\bea{}
{} [f_{a}(\om), f^\dag_{a}(\om')] = C_{aa}\,\delta(\om-\om'),  \quad [f_{b}(\om), f^\dag_{b}(\om')] = C_{bb}\,\delta(\om-\om'), \\
{} [f_{a}(\om), f^\dag_{b}(\om')] = C_{ab}\,\delta(\om-\om'),  \quad [f_{a}(\om), f_{b}(\om')] = D_{ab}\,\delta(\om-\om').
\eea
\esub
 Since $\textrm{det}(H)\ne 0$, \Eq{comms:spdc:abdag}
 reveals that $C_{ab}=0$. The first three equations are four equations in the four (real) unknowns
 $C_{aa}$, $C_{bb}$, $\textrm{Re}(D_{ab})$, $\textrm{Im}(D_{ab})$ which therefore have a unique solution given by \cite{Alsing_Hach:2017a}
\bsub
 \bea{solns:comms}
 C_{kk}(\om) &=& 1 - \alpha_k^2 - |r_k|^2 = 1 -  e^{-\gamma'_k T_k} - |g\alpha_p\,T_k|^2  \myover{{\longrightarrow} {\textrm{high Q}}} \gamma'_k\,T_k - |g\alpha_p\,T_k|^2, \quad k\in\{a,b\}, \qquad\\
 D_{ab} &=& i\, (r^*_b - r_a) = i\,g\,\alpha_p\,(T_b - T_a), \label{solns:comms:Dab}
 \eea
\esub
where  for $C_{aa}(\om)$ and $C_{bb}(\om)$ we have also indicated their values in the high cavity Q limit.
Note, the high cavity Q limit is
defined through the physical conditions (see \cite{Raymer:2013}, Section III)
(i) the cross coupling $\tau_a$ is very small so that the cavity storage time is long,
(ii) the cavity round trip time $T_a$ is small compared to the duration of the input-field pulse i.e. $\om\,T_a \ll 1$, and
(iii) the input field is narrow band and thus well contained within a single FSR of the mrr.
Therefore, with the inclusion of internal propagation loss, the high cavity Q limit is defined by the limits
\bea{RM:HCQ:limit}
\rho_a & \equiv & e^{-\g_a T_a/2} \approx 1-\g_a T_a/2,
\qquad \tau_a = \sqrt{1-\rho_a^2} \approx  \sqrt{\g_a T_a}, \no
 \alpha_a &=& e^{-\g'_a T_a/2} \approx 1-\g'_a T_a/2 ,
\qquad  e^{i\,\om\,T_a} \approx 1 + i\,\om\,T_a.
\eea


\section{The squeezed vacuum state annihilated by $\mathbf{a_{out}(\om)}$ and $\mathbf{b_{out}(\om)}$ and the unitary evolution operator}\label{sec:squeezed:state}
\subsection{Squeezed vacuum state}\label{sec:squeezed:state:vacuum}
In this section we consider the form of the squeezed vacuum state $\ket{0}_{out}$ annihilated by the output
operators $a_{out}(\om)$ and $b_{out}(\om)$ from a Heisenberg  operator perspective.
An expression for $\ket{0}_{out}$ is needed for example when
one computes output correlation functions using the input operators
(employed say in formulating input states) expressed in terms of the
output and noise operators via the inversion of  \Eq{aout:matrix:form},
in the form $a_{in}(\om) = G^{-1}(\om)\,[\vec{a}_{out}(\om) - H(\om)\, \vec{f}(\om)]$.

The "input" vacuum $\ket{0}_{in}\equiv\ket{0}_a\,\ket{0}_b\,\ket{vac}_{env}$ for the signal and idler modes $a,\,b$ and the noise (\textit{environment}) modes $\tf_a,\,\tf_b$
at the entrance port to the mrr is the usual vacuum annihilated by the input operators, namely
$a_{in}\,\ket{0}_a=b_{in}\ket{0}_b=\tf_a\,\ket{vac}_{env}=\tf_b\,\ket{vac}_{env}=0$.
After the process of pair production within the mrr, the Heisenberg operators at the input port of the mrr are unitarily transformed from $a_{in}\rightarrow a_{out}$ at the output port as given by \Eq{aout:matrix:form}.
Equivalently, one can consider the vacuum state as being transformed from
$\ket{0}_{in}\rightarrow\ket{0}_{out}$, where the "out" vacuum is defined as that state annihilated by the output operators, i.e. $a_{out}\,\ket{0}_{out}=b_{out}\ket{0}_{out}=0$.
In the absence of loss and noise, the unitary operator that affects this transformation is the two-mode squeezing operator \cite{Walls_Milburn:1994,Gerry_Knight:2004,Agarwal:2013}
$U_{out,in}=S(\xi) = \exp[\half\,(\xi\,\a^\dag_{in}\,b^\dag_{in} - \xi^*\,a{in}\,b_{in})]$ such that the two-mode squeezed vacuum is given by
$\ket{0}_{out} = S(\xi)\ket{0}_{in} = \cosh^{-1} r\,\sum_{n=0}^\infty\,\tanh^nr\,\ket{0}_a\,\ket{0}_b$ where
$\xi = r\,e^{i\phi}$ is the complex squeezing parameter of magnitude $r$. Here, we wish to find the appropriate squeezed vacuum state $\ket{0}_{out}$ and unitary transformation $U$ when we allow for loss, and retain the noise operators $\tf_a,\,\tf_b$ in $a_{out}$ and $b_{out}$ as in \Eq{aout:matrix:form}.

From \Eq{aout:matrix:form} we wish to solve the operator equations (dropping the argument $\om$ in this section)
\bsub
\bea{a:out:eqn}
a_{out}\ket{0}_{out} &=&
\left[G_{aa}\,a_{in} + G_{ab}\,b^\dag_{in} + \tH_{aa}\,\tf_a + \tH_{ab}\,\tf^\dag_b\right]\ket{0}_{out}=0, \\
b_{out}\ket{0}_{out} &=&
\left[ G^*_{ba}\,a^\dag_{in} + G^*_{bb}\,b_{in} + \tH^*_{ba}\,\tf^\dag_a + \tH^*_{bb}\,\tf_b \right]\ket{0}_{out}=0, \label{b:out:eqn}
\eea
\esub
where we have rescaled $H$ and $f$ such that $H\,f = \tH\,\tf$
where $\tH_{ij} = H_{ij}\,\sqrt{C_{jj}}$ and $\tf_{j} = f_{j}/\sqrt{C_{jj}}$ so that $[\tf_i,\tf^\dag_i]=1$.
In addition, we consider the physically relevant case where $T_a=T_b=T$ so that by \Eq{solns:comms:Dab}
we have $D_{ab}=0$ so that $[\tf_i,\tf^\dag_j]=\delta_{ij}$, implying now that the $in$ and noise operators
$a_{in}$, $b_{in}$, $\tf_a$, and $\tf_b$, \textit{all} mutually commute with each other.
(Note that $C_{jj}$ still contains power dependent contributions).

To solve \Eq{a:out:eqn} and \Eq{b:out:eqn} we seek a solution of the form
\be{S}
S =
\exp\left[
   A\,a^\dag_{in}\,b^\dag_{in} + B\,\tf_a^\dag\,\tf_b^\dag
+  A^{'}\,a^\dag_{in}\,\tf_b^\dag + B^{'}\,\tf^\dag_a\,b^\dag_{in}\right]\ket{0}_{in}
\equiv
\exp[ \mathcal{O}^\dag ]\ket{0}_{in}.
\ee
We note that for any annihilation operator $a$ such that $[a, a^\dag]=1$ and $a\,\ket{0}_{in}=0$,
and for any operator function $f(a, a^\dag)$ one has
$a\,f(a, a^\dag)\,\ket{0}_{in} = [a, f(a, a^\dag)]\ket{0}_{in} = \partial_{a^\dag}\,f(a, a^\dag)\ket{0}_{in}$.
For the form of $S$ given in \Eq{S} this implies
$a_{in}\,S\,\ket{0}_{in} = (\partial_{a^\dag_{in}}\,\mathcal{O}^\dag)\,S\,\ket{0}_{in}$
leading to the operator equations
\bsub
\bea{opr:eqns}
\left[G_{aa}\,(\partial_{a^\dag_{in}}\,\mathcal{O}^\dag)  + G_{ab}\,b^\dag_{in}
    + \tH_{aa}\,(\partial_{\tf_a^\dag}\,\mathcal{O}^\dag) + \tH_{ab}\,\tf^\dag_b\right] S\ket{0}_{in}&=&0,\\
\left[ G^*_{ba}\,a^\dag_{in}  + G^*_{bb}\,(\partial_{b^\dag_{in}}\,\mathcal{O}^\dag)
     + \tH^*_{ba}\,\tf^\dag_a + \tH^*_{bb}\,(\partial_{\tf_b^\dag}\,\mathcal{O}^\dag) \right] S\ket{0}_{in}&=&0.
\eea
\esub
Using the explicit expression for $\mathcal{O}^\dag$ in \Eq{S}, and equating the resulting  coefficients of
the operators $a^\dag_{in}$, $b^\dag_{in}$, $f^\dag_a$ and $f^\dag_b$ to zero leads
to a set of four linear equations
\bsub
\bea{opr:eqns:all}
G_{aa}\,A       + \tH_{aa}\,B^{'}   &=& -G_{ab}, \label{opr:eqn:1}\\
G^*_{bb}\,A     + \tH^*_{bb}\,A^{'} &=& -G^*_{ba}, \label{opr:eqn:2}\\
G_{aa}\,A^{'}   + \tH_{aa}\,B       &=& -\tilde{H}_{ab}, \label{opr:eqn:3}\\
G^*_{bb}\,B^{'} + \tH^*_{bb}\,B     &=& -\tilde{H}^*_{ba}. \label{opr:eqn:4}
\eea
\esub
The determinant of the above four linear equations for the four unknown coefficients $A, B, A', B'$ is zero, indicating that there are only three independent equations. Using the \Eq{comms:spdc:ab} for the condition $[a_{out}, b_{out}] = 0$, written in the form
$-G_{ab}\,G^*_{bb} + \tH_{aa} \, \tH^*_{ba} = -G_{aa}\,G^*_{ba} + \tH_{ab}\,\tH^*_{bb}$,
we observe that upon solving for $A'$ in \Eq{opr:eqn:3} and $B'$ in \Eq{opr:eqn:4} in terms of $B$,
and substituting into \Eq{opr:eqn:1} and \Eq{opr:eqn:2}, $A$ can be written in two equivalent forms
\be{A:soln}
A = \fracd{-G_{ab}\,G^*_{bb} + \tH_{aa}\,\tH^*_{ba}}{G_{aa}\,G^*_{bb}}
     + \fracd{\tH_{aa}\,\tH^*_{bb}}{G_{aa}\,G^*_{bb}}\,B
= \fracd{-G_{aa}\,G^*_{ba} + \tH_{ab}\,\tH^*_{bb}}{G_{aa}\,G^*_{bb}}
     + \fracd{\tH_{aa}\,\tH^*_{bb}}{G_{aa}\,G^*_{bb}}\,B.
\ee
Since the coefficient multiplying $B$ is identical in both terms in \Eq{A:soln}, this implies $B$ is an undetermined free parameter. Here, we take $B=0$ as the simplest choice so that the coefficients in the operator
$S = e^{\mathcal{O}^\dag}$ in \Eq{S} are given by
\be{A:Apr:Bpr:solns}
A =
  -\fracd{G_{ab}}{G_{aa}}
+ \fracd{\tH_{aa}\,\tH^*_{ba}}{G_{aa}\,G^*_{bb}}
=
  -\fracd{G^*_{ba}}{G^*_{bb}}
+ \fracd{\tH_{ab}\,\tH^*_{bb}}{G_{aa}\,G^*_{bb}}, \qquad
A' = -\fracd{\tH_{ab}}{G_{aa}}, \qquad
B' = -\fracd{\tH^*_{ba}}{G^*_{bb}},
\ee
containing a signal-idler pair production $A$, an idler loss term $A'$, and a signal loss term $B'$. In general, a non-zero $B$ term would contribute to corrections to the bare vacuum $\ket{0}_{in}$.

\subsection{Unitary evolution operator}\label{sec:squeezed:state:U}
To construct an evolution operator $U$ such that
$\vec{a}_{out}=U\,\vec{a}_{in}\,U^\dag$ as per \Eq{aout:matrix:form},
we note that the two-mode squeezing operator
$U_{\tA} = \exp[\tA\,a^\dag_{in}\,b^\dag_{in} - \tA^*\,a_{in}\,b_{in}]$ transforms
$U_{\tA}\,a_{in}\,U^\dag_{\tA} = \cosh |\tA|\,a_{in} - e^{i\theta_{\tA}}\,\sinh |\tA|\,b^\dag_{in}$ and
$U_{\tA}\,b^\dag_{in}\,U^\dag_{\tA} = \cosh |\tA|\,b^\dag_{in} - e^{-i\theta_{\tA}}\,\sinh |\tA|\,a_{in}$
where
$\tA = |\tA|\,e^{i\,\theta_{\tA}}$.
Let us also similarly define
$U_{\tA'} = \exp[\tA'\,a^\dag_{in}\,\tf^\dag_b - \tA^{'*}\,a_{in}\,\tf_b]$
and
$U_{\tB'} = \exp[\tB'\,\tf^\dag_a\,b^\dag_{in} - \tB^{'*}\,\tf_b\,b_{in}]$
with $\tA' = |\tA'|\,e^{i\,\theta_{\tA'}}$ and
$\tB' = |\tB'|\,e^{i\,\theta_{\tB'}}$,
and lastly
$U_{\theta} =\exp[-i\,\theta_{G_{aa}}\,a^\dag_{in}\,a_{in} + i\,\theta_{G_{bb}}\,b^\dag_{in}\,b_{in} ]$ where $G_{aa} = |G_{aa}|\,e^{i\,\theta_{G_{aa}}}$ and
 $G_{bb} = |G_{bb}|\,e^{i\,\theta_{G_{bb}}}$.
Then the operator
\be{U}
U = U_{\tB'}\,U_{\tA'}\,U_{\tA}\,U_{\theta}
\ee
implements the transformation
\be{U:transform}
\left[
  \begin{array}{c}
    a_{out} \\
    b^\dag_{out} \\
  \end{array}
\right]
=
U\,
\left[
  \begin{array}{c}
    a_{in} \\
    b^\dag_{in} \\
  \end{array}
\right]
\,U^\dag
=
\left[
  \begin{array}{c}
    G_{aa}\,a_{in} + G_{ab}\,b^\dag_{in} + \tH_{aa}\tf_a + \tH_{ab}\,\tf^\dag_b \\
    G_{ba}\,a_{in} + G_{bb}\,b^\dag_{in} + \tH_{ba}\tf_a + \tH_{bb}\,\tf^\dag_b  \\
  \end{array}
\right],
\ee
with the assignments (after some straightforward algebra)
\be{amplitudes:phases}
\begin{array}{cclccl}
G_{aa} &=& \cosh|\tA|\,\cosh|\tA'| e^{i\,\thaa}, &
G_{ba} &=& -\sinh|\tA|\,\cosh|\tA'|\,e^{i\,(\theta_{\tA} + \thbb)}, \\
G_{ab} &=& -\sinh|\tA|\,\cosh|\tB'|\,e^{i\,(\theta_{\tA} + \thaa)}, &
G_{bb} &=&  \cosh|\tA|\,\cosh|\tB'|\,e^{i\,\thbb},\\
\tH_{aa} &=&   \sinh|\tA|\,\sinh|\tB'|\,e^{i\,(\theta_{A} -\theta_{\tB'}+\thaa)}, &
\tH_{ba} &=&  -\cosh|\tA|\,\sinh|\tB'|\,e^{i\,(-\theta_{\tB'}+\thbb)},\\
\tH_{ab} &=& \cosh|\tA|\,\sinh|\tA'|\,e^{i\,(\theta_{\tA'}+\thaa)}, &
\tH_{bb} &=& \sinh|\tA|\,\sinh|\tA'|\,e^{i\,(-\theta_{\tA} + \theta_{\tA'} + \thbb)}.\\
\end{array}
\ee
These assignments identically satisfy the output
commutator relations in \Eq{comms:spdc:aadag}, \Eq{comms:spdc:bbdag}
and \Eq{comms:spdc:ab} for arbitrary $\tA,\,\tA',\,\tB'$.
Substituting \Eq{amplitudes:phases} into \Eq{A:Apr:Bpr:solns}
yields the identifications
\be{A:Apr:Bpr:v2}
\begin{array}{ccl}
|A| &=& \fracd{|G_{ab}|\,|G_{bb}| - |\tH_{aa}|\,|\tH_{ba}|}{|G_{aa}|\,|G_{bb}|} =
\fracd{\tanh|\tA|}{\cosh|\tA'|\,\cosh|\tB'|}, \\
|A'| &=& \fracd{|\tH_{ab}|}{|G_{aa}|} = \tanh|\tA'|\\
|B'| &=& \fracd{|\tH_{ba}|}{|G_{bb}|} = \tanh|\tB'|, \\
\end{array}
\ee
where the phases $e^{i\,\theta_{\tX}}$ for $\tX\in\{\tA, \tA', \tB'\}$ have identically canceled on both sides of the equalities in the three formulas in \Eq{A:Apr:Bpr:solns}
if we take
$\theta_{\tA}=\theta_{A}$,
$\theta_{\tA'}=\theta_{A'}$, and
$\theta_{\tB'}=\theta_{B'}$.

Note that in the weak field limit $|g\,\alpha_p\,T|\ll 1$
we have $|A|, |A'|, |B'|\sim \mathcal{O}(|g\,\alpha_p\,T|)$ since each contains an off-diagonal element of either the $G$ or $\tH$ matrices.
Assuming the same holds true for $\tX\in\{\tA,\,\tA',\,\tB'\}$ justifies the use of the first order approximations
$\tanh|\tX| \approx |\tX|$ and $\cosh|\tX|\approx 1$.
Under these conditions
the three equations in \Eq{A:Apr:Bpr:v2} simply reduce to $|\tA|\approx |A|$,
$|\tA'|\approx |A'|$, and  $|\tB'|\approx |B'|$ which are effectively what has been utilized in the previous sections to produce the two-photon state.
Lastly, note that without the transformation $U_\theta$ in \Eq{U} the quantities $G_{aa}$ and $G_{bb}$ would have been assigned real values in \Eq{amplitudes:phases} under the remaining transformations alone. Thus, $U_{\theta}$ was introduced to take into account the complexity of $G_{aa}$ and $G_{bb}$ by introducing the phases $e^{i\,\thaa}$ and  $e^{i\,\thbb}$ in \Eq{amplitudes:phases}.

\subsection{Entanglement in two-photon mixed output state}\label{sec:squeezed:state:Ent}
In this section we compute the entanglement between the generated signal and idler modes of the \textit{output} mixed Gaussian two-photon state in the presence of loss.
For the entanglement measure we compute the log negativity \cite{Adesso:2004,Plenio:2005} (see also \cite{Pirandola_Lloyd:2008}, and pp. 66-67 of \cite{Agarwal:2013} for succinct reviews). The log negativity $E_N(\rho)$ for a mixed Gaussian state $\rho$ is given by $E_N(\rho) = \textrm{max}[0,-\ln(2\tnu_<)]$ where $\tnu_<$ is the smaller of two symplectic eigenvalues $\tnu_\pm$ of the real, positive, symmetric covariance matrix $\sigma_{ij}$,
\be{sigma:defn}
\sigma_{ij} =
\half
\langle
X_i\,X_j + X_j\,X_i
\rangle
-
\langle X_i\rangle\,\langle X_j\rangle,
\ee
which defines the Gaussian mixed state.
In the above, $X_i = (x_a, y_a, x_b, y_b)$  is the row vector of  quadrature variables where
$x_a = (a + a^\dag)/\sqrt{2}$, $y_a = (a - a^\dag)/(\sqrt{2}\,i)$,
$x_b = (b + b^\dag)/\sqrt{2}$, $y_b = (b - b^\dag)/(\sqrt{2}\,i)$,
such that the Wigner function for the normalized Gaussian state is given by
$W(X) = \exp[-(X-\langle X\rangle)\,\sigma^{-1}\,(X-\langle X\rangle)^{T}]/[(2\,\pi)^n\,\sqrt{\textrm{det}(\sigma)}]$ \cite{Agarwal:2013}.
Entanglement is present in the state when $\tnu_< < \half$, yielding $E_N(\rho)~>~0$.

The log negativity capitalizes upon the symplectic structure of the Gaussian correlation matrix. For Gaussian states, linear optical operations simply transform the covariance matrix $\sigma$, while retaining the Gaussian structure of the transformed state. Under linear optical transformations it becomes relatively straightforward to compute bounds on the discrimination of different transformed Gaussian states \cite{Pirandola_Lloyd:2008}. This advantage of quantifying Gaussian states has currently found great utility in analyzing the security of QKD systems based on quantum illumination \cite{Shapiro:2016}, and for the development schemes to detect the residual correlations \cite{Guha:2009} between the interrogating signal and (memory) held idler of the two-mode squeezed state used to determine the presence or absence of a remote target. With the respect to the work investigated here, a mrr is essentially a linear optical beam splitter with passive feedback, whose transformation properties preserve the Gaussian nature of the two-mode squeezed state in the presence of loss.

In \Eq{sigma:defn} above we take expectation values of \textit{in} operators with respect to the \textit{out} state $\ket{0}_{out}$.
Thus,
$\Tr_{sys,env}[\ket{0}_{out}\bra{0}\,\mathcal{G}(\vec{a}_{in},\vec{a}^\dag_{in})]=
{}_{out}\bra{0}\,\mathcal{G}(\vec{a}_{in},\vec{a}^\dag_{in})\,\ket{0}_{out}
={}_{in}\bra{0}\,U^\dag\,\mathcal{G}(\vec{a}_{in},\vec{a}^\dag_{in})\,U\,\ket{0}_{in}
={}_{in}\bra{0}\,\mathcal{G}(\vec{a}_{out},\vec{a}^\dag_{out})\,\ket{0}_{in}$
where $\mathcal{G}(\vec{a}_{in},\vec{a}^\dag_{in})$ is some function of the input operators
\footnote{
Note: Since $U^\dag=e^{-(\mathcal{O}^\dag-\mathcal{O})}=U^{-1}$, the transformation
$U^\dag\,\mathcal{G}(\vec{a}_{in},\vec{a}^\dag_{in})\,U$ actually  transforms the \textit{in} operators with
$U^{-1}(A,A',B') = U(-A,-A',-B')$ with $A, A', B'$ from the previous section.
This is equivalent to $\theta_j\rightarrow \theta_j +\pi$ for $j\in\{A, A', B'\}$
which from \Eq{amplitudes:phases} changes $Z_{ab}\rightarrow -Z_{ab}$ and $Z_{ba}\rightarrow -Z_{ba}$, but
leaves $Z_{aa}$ and  $Z_{bb}$ invariant for $Z\in\{G, \tH\}$. This will will induce
$\mB\rightarrow -\mB$ and $\mC\rightarrow -\mC$ in \Eq{sigma:xaxb} and \Eq{sigma:xayb},
but leave $\mA$ and  $\mA'$ invariant.
However, the symplectic eigenvalues $\tnu_{\pm}$ depend only on $\mB^2, \mC^2$
in \Eq{nupm}, and hence are invariant to this sign change.
Therefore, for consistency of notation and phase assignments, we will compute quadrature amplitudes $\sigma_{ij}$ in \Eq{sigma:matrix} with
$\vec{a}_{out} = U\,\vec{a}_{in}\,U^\dag=G(\om)\,\vec{a}_{in} + \tH(\om)\,\vec{\tf}$ in the following.
},
and the trace is take over both the system (\textit{sys}) $a, b$ and  environment (\textit{env}) $\tf_a, \,\tf_b$ subsystems.
Thus, defining $x_a\rightarrow (a_{out} + a^\dag_{out})/\sqrt{2}$ etc.,  we find using
\bsub
\bea{a:out:defn}
a_{out} &=& G_{aa}\,a_{in} + G_{ab}\,b^\dag_{in} + \tH_{aa}\,\tf_a + \tH_{ab}\,\tf^\dag_b, \\
b^\dag_{out} &=& G_{ba}\,a_{in} + G_{bb}\,b^\dag_{in} + \tH_{ba}\,\tf_a + \tH_{bb}\,\tf^\dag_b, \label{b:out:defn}
\eea
\esub
that the covariance matrix has the form
\be{sigma:matrix}
\sigma =
\left(
  \begin{array}{cccc}
     \mA &   0 &  \mB  & \mC \\
      0  & \mA & -\mC  & \mB \\
    \mB  & \mC &  \mA' &   0 \\
   -\mC  & \mB &  0    & \mA' \\
  \end{array}
\right) \equiv \sigma_G + \sigma_H,
\ee
where
\bsub
\bea{sigma:coeffs}
\mA &=& \sigma_{x_a\,x_a}
= \half
\left[
\left( |G_{aa}|^2 + |G_{ab}|^2  \right)
+
\left( |\tH_{aa}|^2 + |\tH_{ab}|^2 \right)
\right] \equiv \mA_G + \mA_H, \label{sigma:xaxa}\\
\mA' &=& \sigma_{x_b\,x_b}
= \half
\left[
\left( |G_{bb}|^2 + |G_{ba}|^2  \right)
+
\left( |\tH_{bb}|^2 + |\tH_{ba}|^2 \right)
\right] \equiv \mA'_G + \mA'_H, \label{sigma:xbxb}\\
\mB &=&\sigma_{x_a\,x_b}
= \textrm{Re}\left( G_{aa}\,G^*_{ba} + \tH_{aa}\,\tH^*_{ba} \right)  \equiv \mB_G + \mB_H, \label{sigma:xaxb}\\
\mC &=&\sigma_{x_a\,y_b}
= \textrm{Im}\left( G_{aa}\,G^*_{ba} + \tH_{aa}\,\tH^*_{ba} \right)  \equiv \mC_G + \mC_H,\label{sigma:xayb}
\eea
\esub
where \Eq{comms:spdc:ab} has been used to simplify  \Eq{sigma:xaxb} and \Eq{sigma:xayb}.
The matrix $\sigma$ in \Eq{sigma:matrix}
of the state $\rho_{out}= \Tr_{env}[\ket{0}_{out}\bra{0}]$
has the form of a mixed thermal two-mode squeezed state,
whose symplectic eiqenvalues for the covariance matrix associated with its "partial transpose"
are given by (see p67 in \cite{Agarwal:2013})
\be{nupm}
\tnu_{\pm} = \half
\left[
(\mA + \mA') \pm \sqrt{(\mA - \mA')^2 + 4\,(\mB^2 + \mC^2)}
\right].
\ee
The log negativity of the squeezed vacuum state $\rho_{out}$ in the presence of loss is then given by
\be{logNeg:nupm}
E_N^{(out)} = \textrm{max}[ 0, -\ln(2\,\tnu_{<}) ], \qquad \tnu_{<}= \textrm{min}(\tnu_+,\tnu_-),
\quad\Rightarrow\quad E_N^{(out)} > 0 \;\; \textrm{for} \;\; \tnu_{<} < \half,
\ee
where entanglement $E_N^{(out)} > 0$ occurs when $\tnu_{<} < \half$.
The influence of loss on the entanglement of the state can be easily identified in \Eq{sigma:xaxa}-\Eq{sigma:xayb} as the terms $H_{ij}(\om)$
which accompany each corresponding classical-like loss (for $\alpha<1$) term $G_{ij}(\om)$, for $i, j\in\{a,b\}$.

\begin{figure}[ht]
\begin{tabular}{cc}
\includegraphics[width=6.0in,height=2.00in]{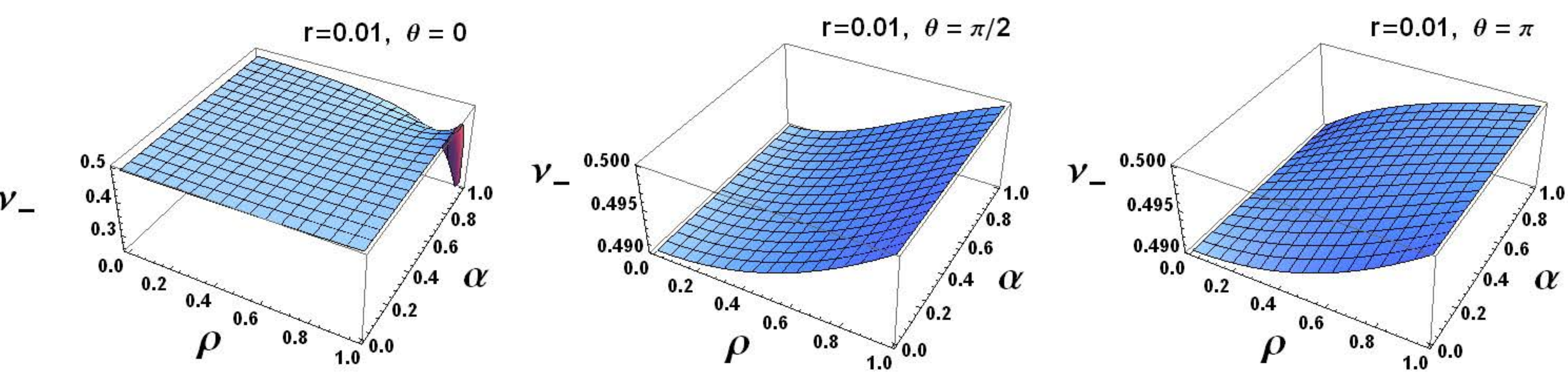}\\
\includegraphics[width=6.0in,height=2.00in]{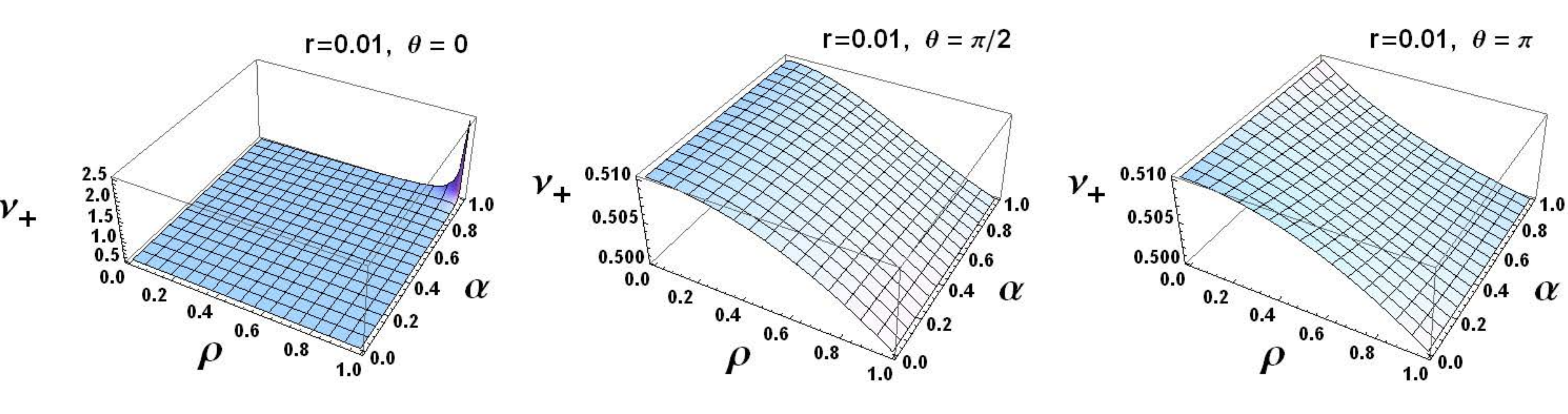}
\end{tabular}
\caption{Symplectic eigenvalues (top) $\tnu_-$ and  (bottom) $\tnu_+$, for (left) phase accumulation angles on mrr resonance $\theta=0$, (middle) slightly off mrr resonance $\theta=\pi/2$, (right) midway between mrr resonances $\theta=\pi$ for the case of signal-idler photon loss.
}\label{fig:nuplus:minus:thetas}
\end{figure}

\begin{figure}[ht]
\begin{tabular}{cc}
\includegraphics[width=3.0in,height=3.5in]{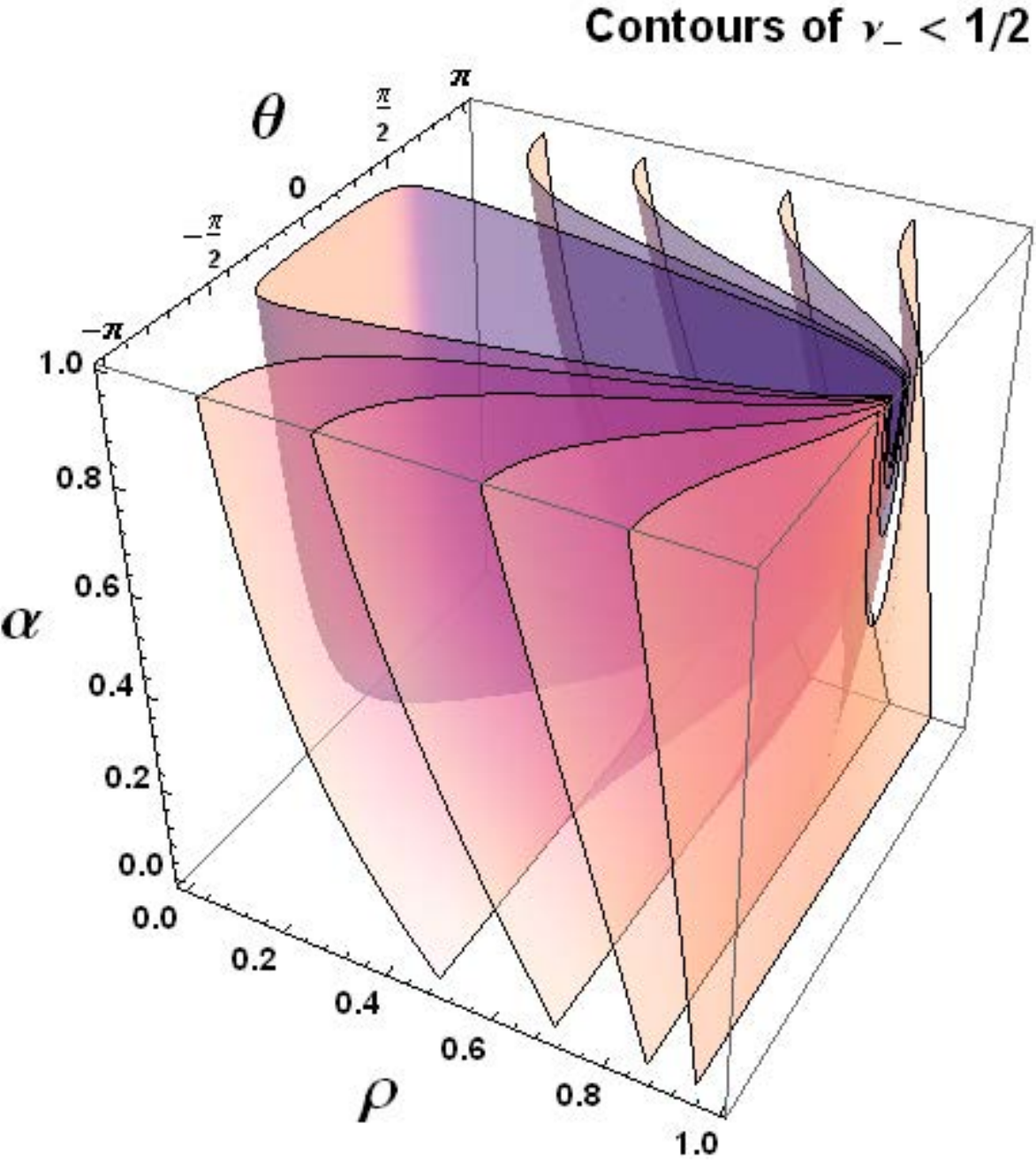} & 
\includegraphics[width=3.0in,height=3.5in]{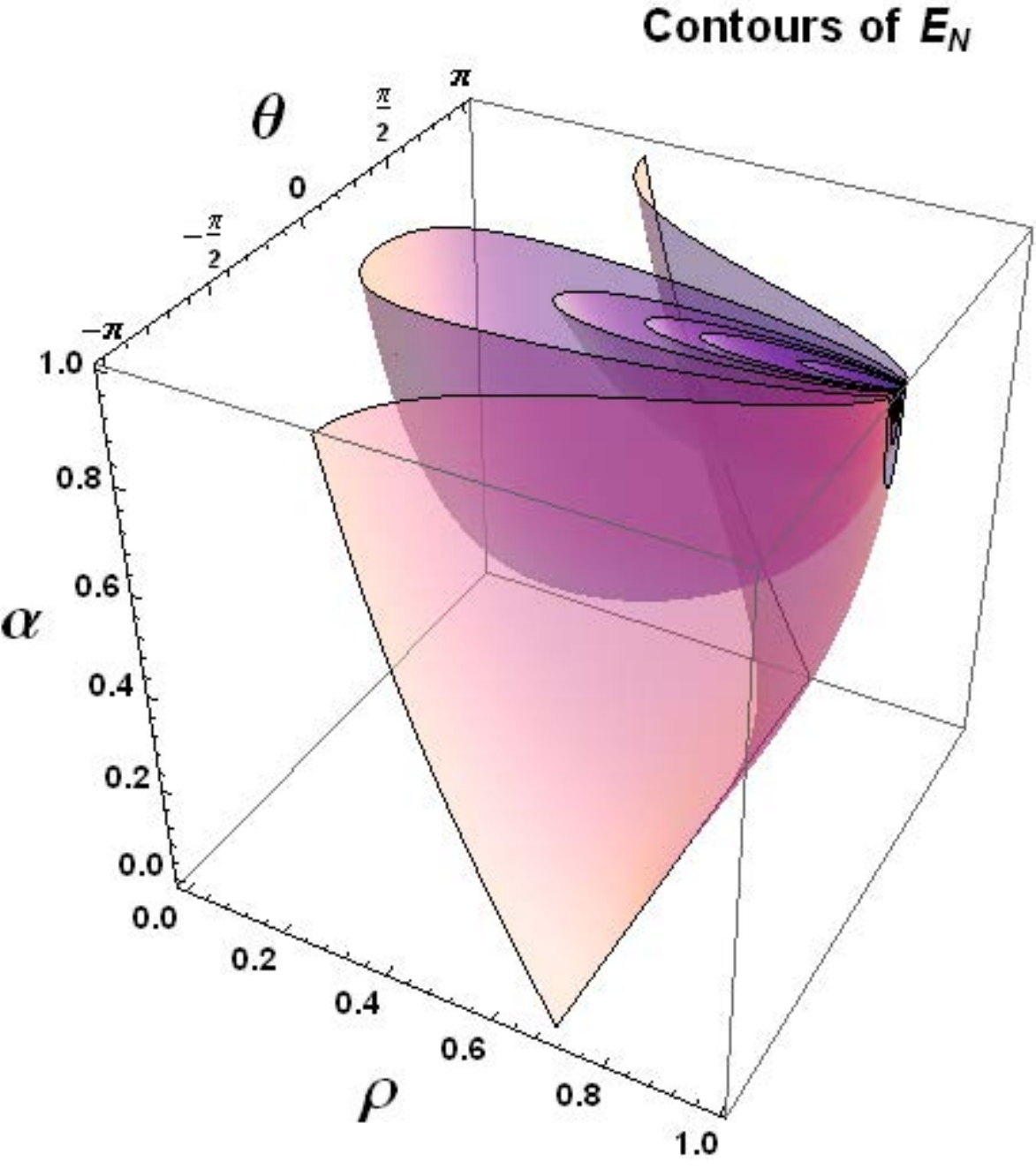}  
\end{tabular}
\caption{Contour plots  of
(left: surfaces from left to  right in the cube) symplectic eigenvalues
$\tnu_{<} = \tnu_{-}\in\{0.490, 0.4925, 0.495, 0.4975, 0.499\}$,
(corresponding to log negativity
$E_N^{(out)}\in\{0.020, 0.015, 0.010, 0.005, 0.002\}$),
and
(right: outer to inner surfaces in cube) log negativity
$E_N^{(out)}\in\{0.01, 0.025, 0.05, 0.075, 0.10, 0.20\}$
(corresponding to symplectic eigenvalues
$\tnu_{-}~\in~\{0.495, 0.488, 0.476, 0.464, 0.452, 0.409\}$) for the case of signal-idler photon loss.
}\label{fig:contours:numinus_EN}
\end{figure}

Let us examine the symplectic eigenvalues for the case of equal self-coupling $\rho_a = \rho_b = \rho$, and equal propagation loss $\alpha_a = \alpha_b = \alpha$ for the signal and idler modes $a$ and $b$, respectively. Since we have considered $T_a=T_b=T$ we also have $\theta_a=\theta_b=\theta=\om\,T$. For simplicity, we take the pump $\alpha_p$ to be real so that $r_a=r_b=g\,\alpha_p\,T\equiv r$.
Since in general $r=g\,\alpha_p\,T\ll 1$ we expand the symplectic eigenvalues to $\mathcal{O}(r^2)$ to obtain
\be{nu:plus:minus}
\tnu_{\pm} \approx
\half
\pm r\,(1-\rho^2)\,|S(\rho,\theta,\alpha)|^2
+ \half\,r^2\,(1-\rho^2)\,|S(\rho,\theta,\alpha)|^4\,(3-\alpha^2\,\rho^2 - 2\,\alpha\,\rho\,\cos\theta),
\ee
where
$|S(\rho,\theta,\alpha)|^2 = 1/|1-\rho\,\alpha\,e^{i\theta}|^{2} = (1+\rho^2\,\alpha^2 - 2\,\rho\,\alpha\,\cos\theta)^{-1}$ is the modulus squared of the round trip circulation factor.
In \Fig{fig:nuplus:minus:thetas} we plot the full expressions for $\tnu_{\pm}$ for which \Eq{nu:plus:minus} is numerically a very good approximation for $r<0.01$.
In general, the mrr has resonances at $\theta=\om\,T=2\,m\,\pi$, and $\theta=(2\,m+1)\,\pi$ represents off-resonance points located midway between cavity resonances (in the middle of the cavity free spectral range).
(Note that experimental values of $r$ for typical pump values of $1$mW are on the order of $r\sim 10^{-5}$ \cite{Schneeloch:2016,Schneeloch:2017}, but in order to illustrate the effects of entanglement we will use a value of $r=0.01$ is the plots below).
Thus we see that  $\tnu_{+} > \half$, and that a small amount of entanglement
occurs whenever  $\tnu_< = \tnu_{-} < \half$.
In \Fig{fig:contours:numinus_EN} we plot the contour values of $\tnu_{-}$ and the corresponding values of the log negativity $E_N^{(out)}(\tnu_-)$ as a function of $0\le\rho,\,\alpha \le 1$ and $-\pi\le\theta\le\pi$ for $r=0.01$.
The more the  symplectic eigenvalue $\tnu_-$ is less than $1/2$, the larger is the log negativity
$E_N^{(out)}(\tnu_-)=-\ln(2\,\tnu_-)$, and hence the larger the entanglement between the generated signal and idler occupation modes (for fixed frequencies that add up to the pump frequency for SPDC, or twice the pump frequency for SFWM). In the \Fig{fig:contours:numinus_EN}(left) we plot the prominent contour values of $\tnu_-$ near $1/2$ since the loss has degraded the entanglement (small values of $E_N^{(out)}(\tnu_-)$). However, in \Fig{fig:contours:numinus_EN}(right) slightly larger values of the log negativity do exist in the presence of loss, however, these contour surfaces (from outer to inner) become smaller as the value of $E_N^{(out)}(\tnu_-)$ increases.
\begin{figure}[h t]
\includegraphics[width=6.0in,height=2.00in]{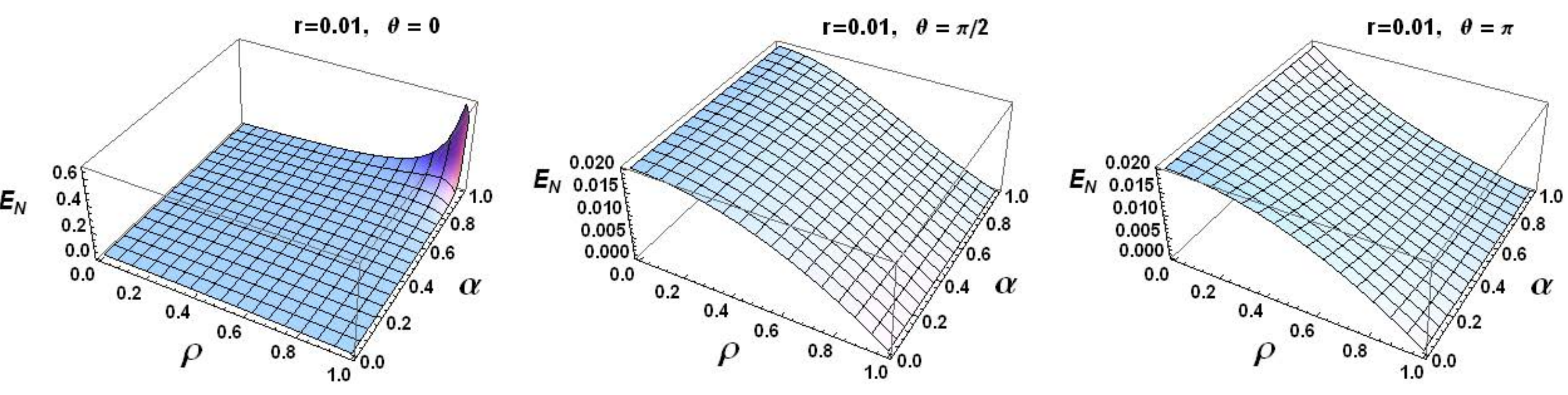}
\caption{$E_N^{(out)}$ for (left) $\theta=0$, (middle) $\theta=\pi/2$, (right) $\theta=\pi$ for the case of signal-idler photon loss, where $\theta = \omega\,T$.
}\label{fig:EN:thetas}
\end{figure}
In \Fig{fig:EN:thetas} we plot the log negativity $E_N^{(out)}(\tnu_-)$ for fixed values of
$\theta = \omega\,T =(0, \pi/2, \pi)$.
We see that that the entanglement is largest for $\theta=0$ which corresponds to resonance condition for the mrr, and drops off precipitously for $\theta$ non-zero. Note that $\theta=\pi$ corresponds to the midpoint in the free spectral range of the mrr between resonances (at integer multiples of $2\pi$).

\begin{figure}[h t]
\begin{tabular}{ccc}
\includegraphics[width=2.5in,height=1.5in]{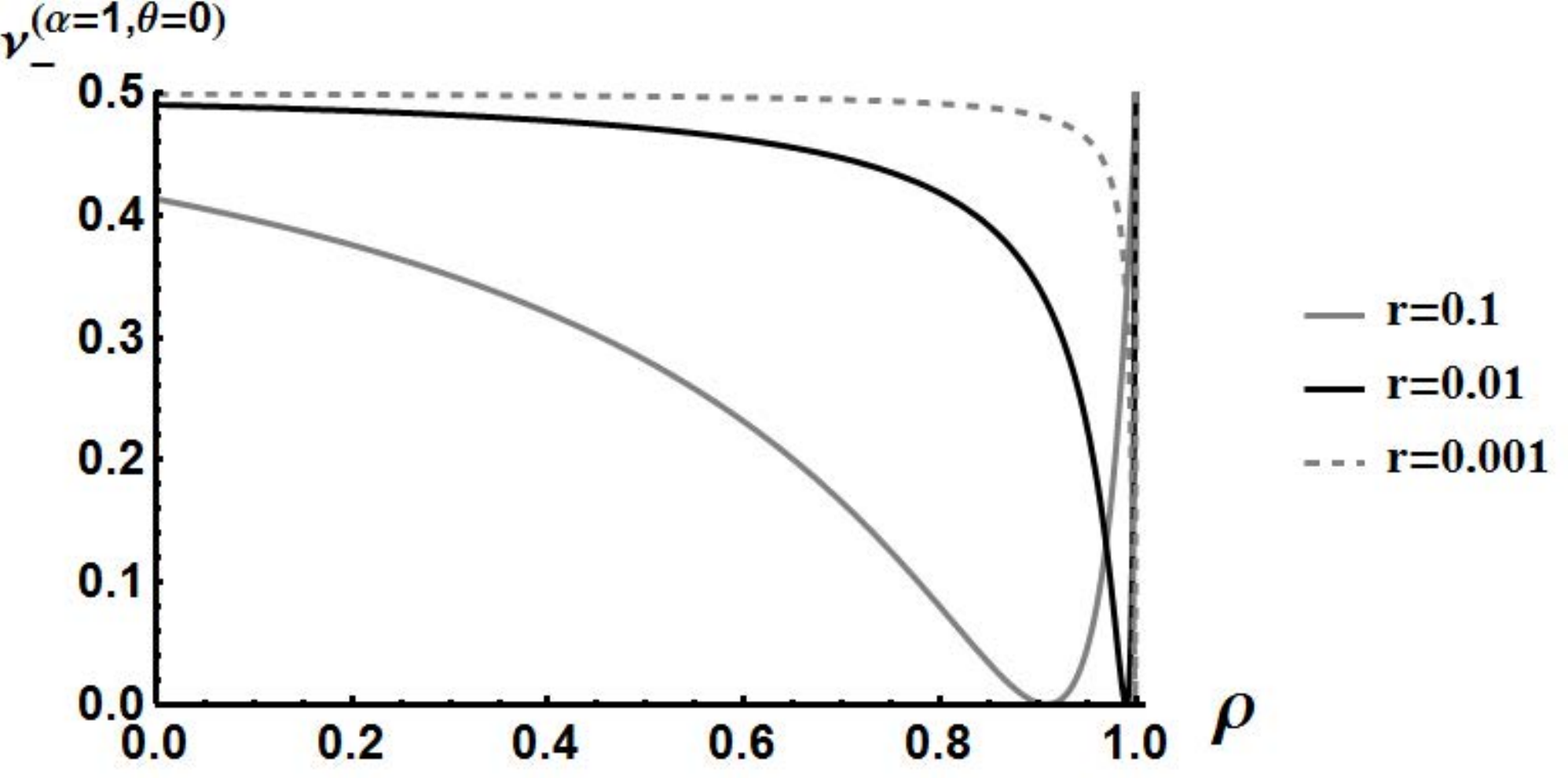} & {\qquad} & 
\includegraphics[width=2.5in,height=1.5in]{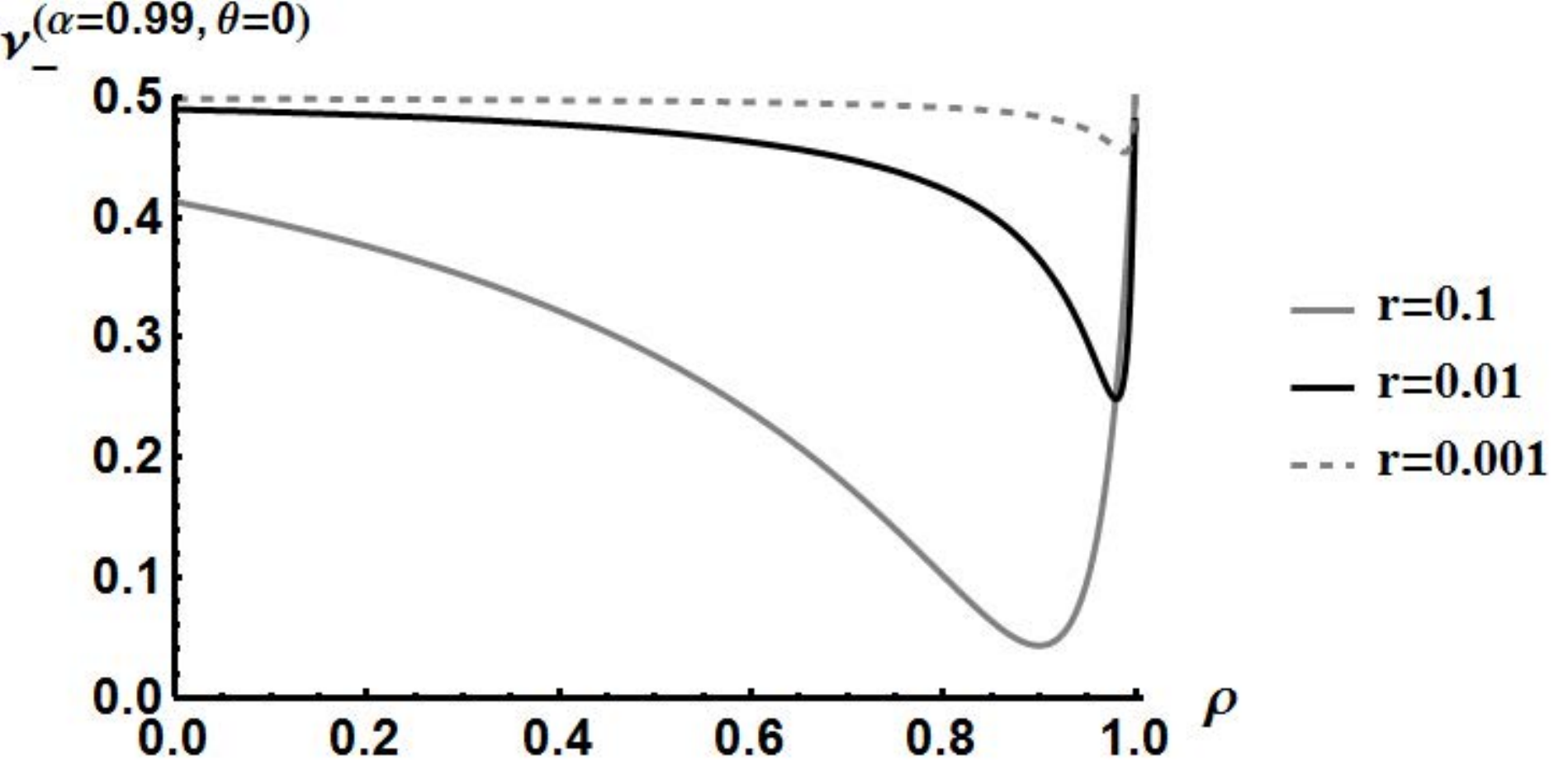} \\ 
\includegraphics[width=2.5in,height=1.5in]{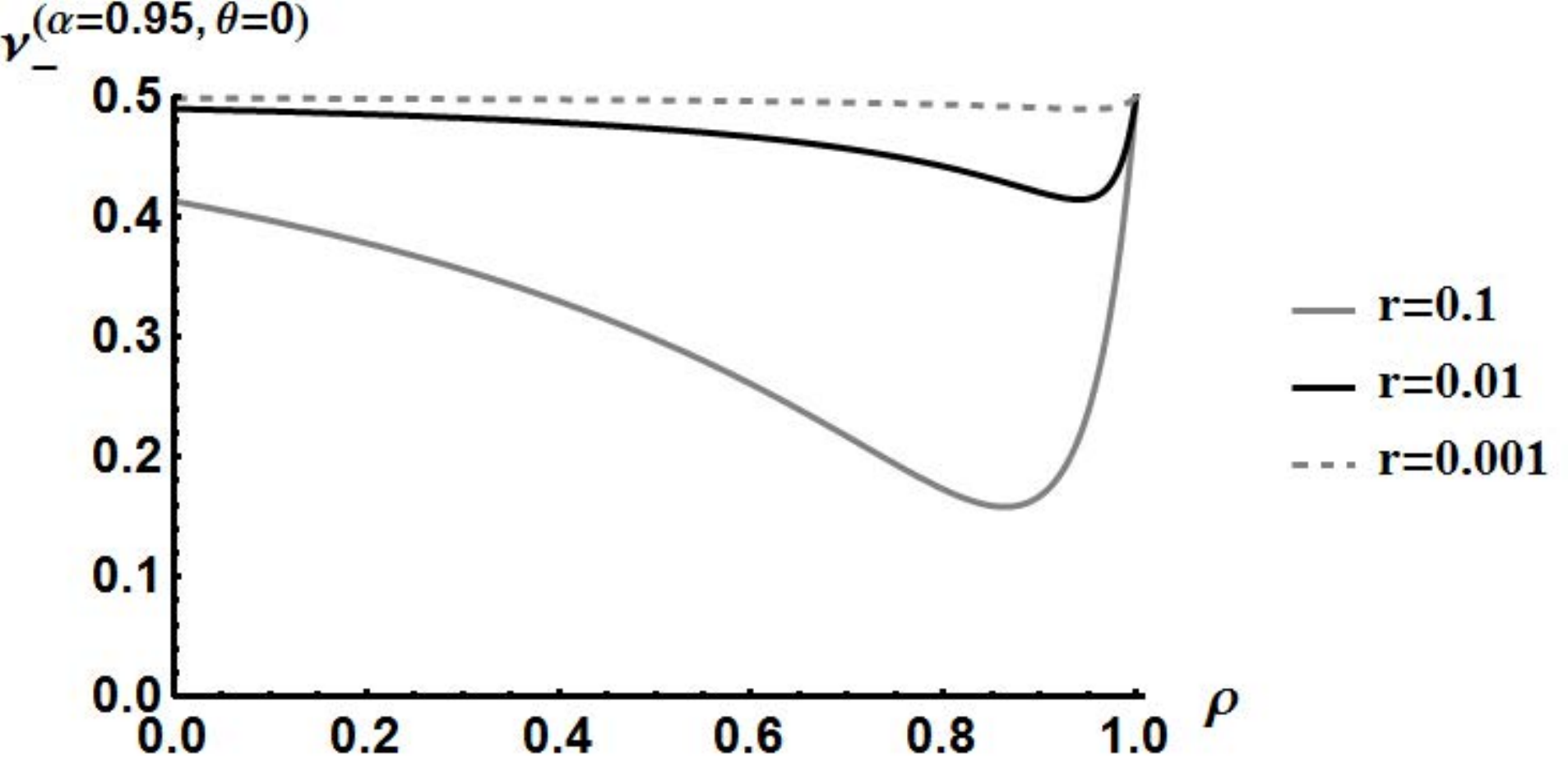} & {\qquad} & 
\includegraphics[width=2.5in,height=1.5in]{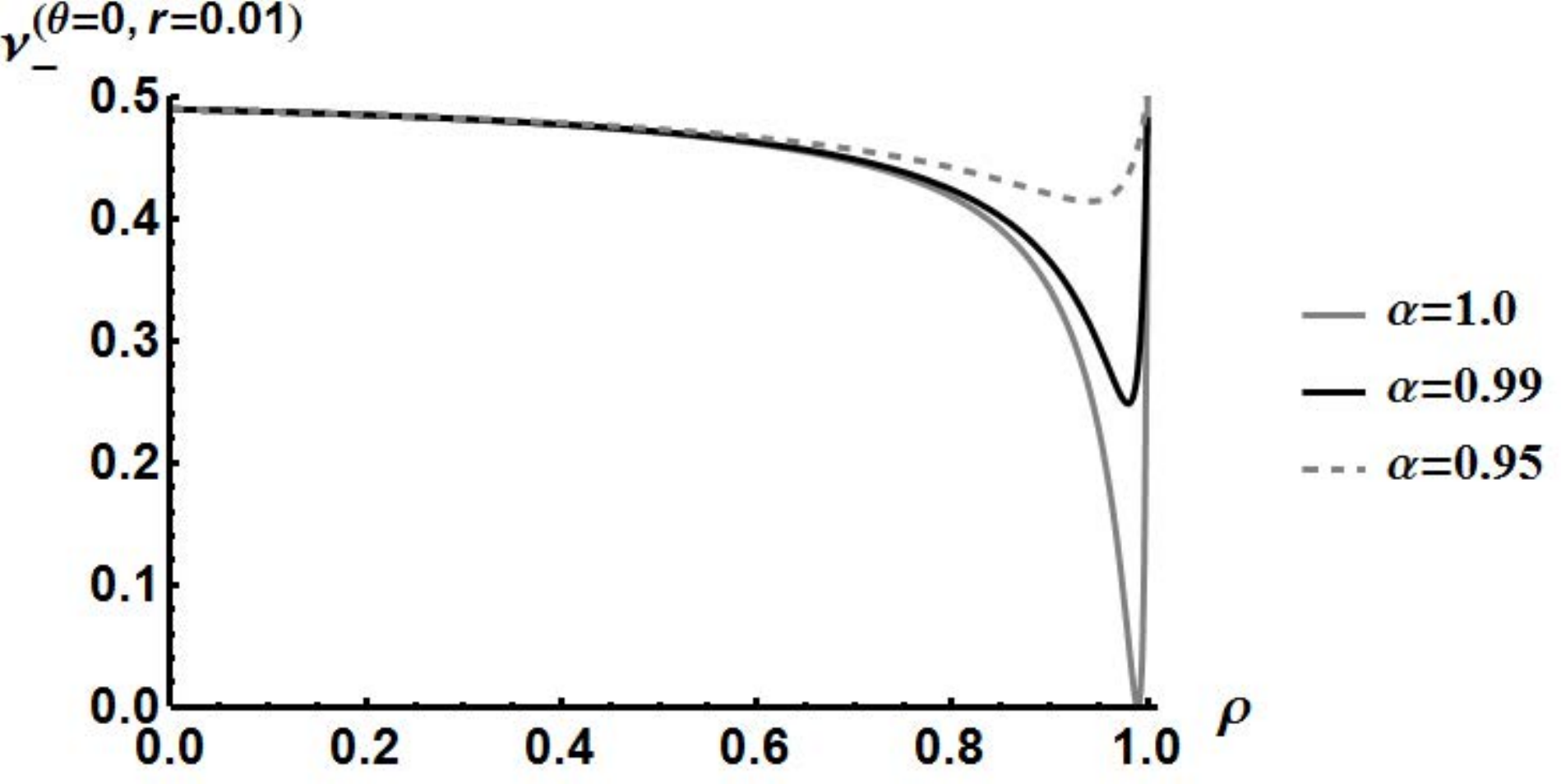} \\ 
\end{tabular}
\caption{$\tnu_{-}^{(\theta=0)}(\rho,r)$ for (upper-left, upper-right, lower-left) $\alpha=(1.0, 0.99, 0.95)$
with (gray-solid) $r=0.1$, (black-solid) $r=0.01$, (gray-dashed) $r=0.001$.
(lower-right) $\tnu_{-}^{(\theta=0, r=0.01)}(\rho)$  with (gray-solid) $\alpha=1.0$, (black-solid) $\alpha=0.99$, (gray-dashed) $\alpha=0.95$.
}\label{fig:numinus:theta0:alphas}
\end{figure}

From the above figures, and from \Eq{nu:plus:minus} we see that as we approach the case of no internal propagation loss in the mrr $\alpha\rightarrow 1$, we have the following limits as we also approach the high Q limit $\rho\rightarrow 1$,
\be{nu:plus:minus:limits}
\tnu_{\pm} =
\half\,
\left\{
\begin{tabular}{c}
$\left( \fracd{1-\rho \pm r\,\rho}{1-\rho \mp r} \right)^2$ \\
$\tf_{\theta=\pi/2}(\rho,r)$ \\
$\left( \fracd{1+\rho \mp r\,\rho}{1+\rho \pm r}\right)^2$
\end{tabular}
\right\}
\rightarrow \half \pm
r\,\left\{
\begin{tabular}{cccll}
$\fracd{1+\rho}{1-\rho}$     &+& $\half\,r\,\fracd{(1-\rho)\,(3+\rho)}{(1-\rho)^2}$ & \quad\textrm{for}  & $\;\theta = 0$, \\
$\fracd{1-\rho^2}{1+\rho^2}$ &+& $\half\,r\,\fracd{(1-\rho^2)\,(3-\rho^2)}{(1+\rho^2)^2}$ & \quad\textrm{for}&$\;\theta =\pi/2$,\\
$\fracd{1-\rho}{1+\rho}$     &+& $\half\,r\,\fracd{(1+\rho)\,(3-\rho)}{(1+\rho)^2}$ & \quad\textrm{for}  & $\;\theta = \pi$
\end{tabular}
\right\},
\ee
where $\tf_{\theta=\pi/2}(\rho,r)$ is an involved function of $\rho$ and $r$ that does not reduce to a simple form for $\theta=\pi/2$.
From \Eq{nu:plus:minus:limits} we can infer that as $r = g\,\alpha_p\,T$  increases (e.g. pump power or cavity round trip time) $\tnu_-$ is driven to be less than a half, thus increasing entanglement, while $\tnu_+$ is driven to be greater than a half. But there is a limit as to how much we can increase say the pump power before other parasitic effects are introduced.
However, by inspection of \Eq{nu:plus:minus:limits}
we can further enhance entanglement on resonance $\theta=0$
so that to $\mathcal{O}(r^2)$ we have $\tnu_- \sim \half - 2\,r/(1-\rho) + 4\,r^2/(1-\rho)^2$ as we approach the high Q limit, thus further decreasing $\tnu_-$ below a half
(for fixed $r<2/3$, where the $\mathcal{O}(r)$ terms equals the $\mathcal{O}(r^2)$ in this approximation).
In \Fig{fig:numinus:theta0:alphas} we plot the full expression $\tnu_{-}^{(\theta=0,\alpha)}(\rho,r)$ as a function of $\rho$ for $r=(0.1, 0.01, 0.001)$
for the ideal lossless case (upper-left) $\alpha=1.0$, showing $\tnu_-(\rho^*)=0$ at $\rho^* = (0.909091, 0.990098, 0.998998)$, respectively.
At such values of $\rho^*$ the squeezing if formally infinite, but this idealized case of no internal propagation loss $\alpha=1$ is used to illustrate the effect of a high cavity Q on the entanglement.
(Note that for the other values of $\theta=(\pi/2, \pi)$ in \Eq{nu:plus:minus:limits},
$\tnu_-\rightarrow \half$ as $\rho\rightarrow 1$).
Of course, the smaller the value of $r$ the closer $\rho^*\rightarrow 1$.

\begin{figure}[h t]
\includegraphics[width=3.5in,height=2.75in]{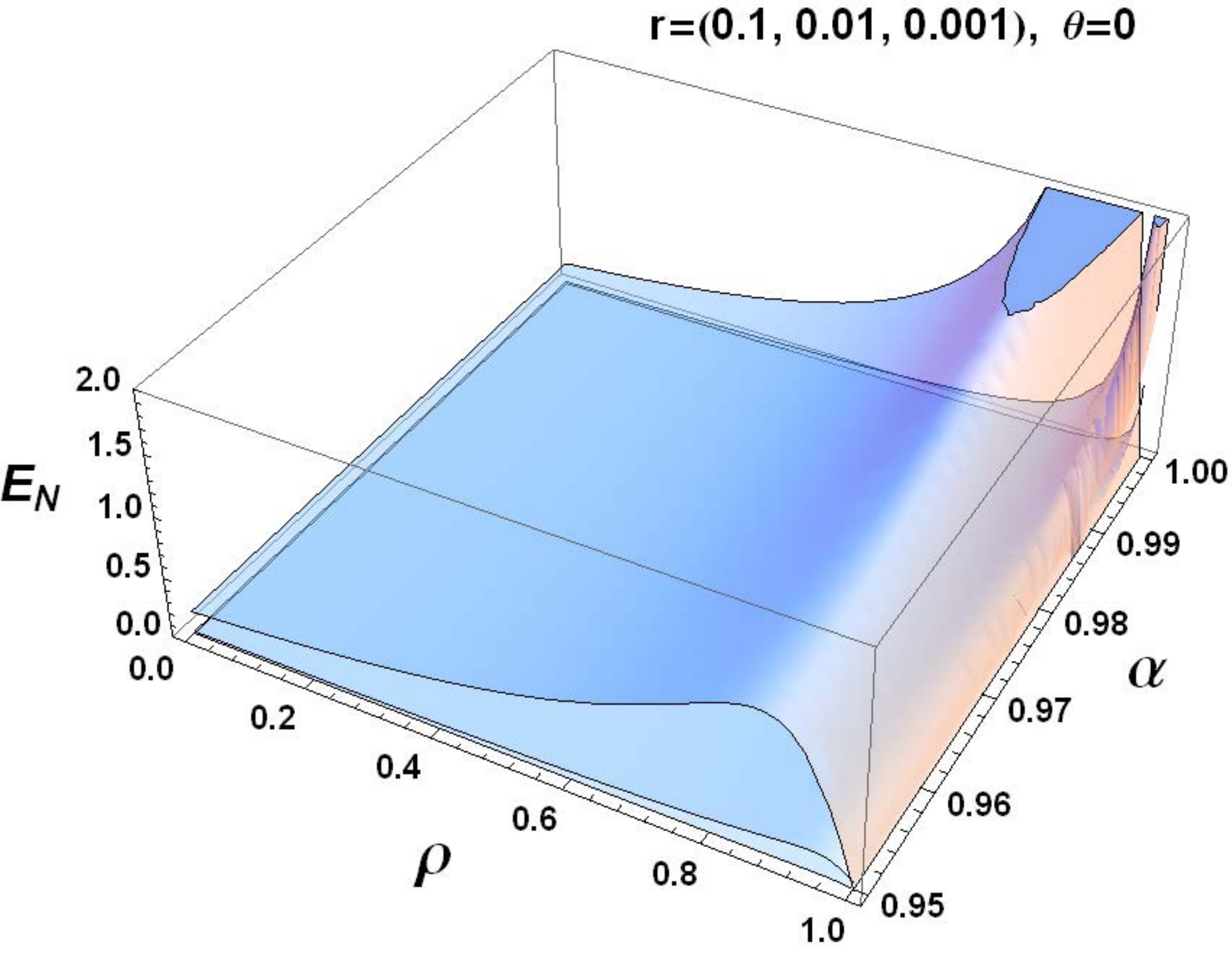} 
\caption{$E_N^{(\theta=0)}(\rho,\alpha)=-\ln[2\,\tnu_{-}^{(\theta=0)}(\rho,\alpha)]$
for (top surface) $r=0.1$, (middle surface) $r=0.01$, (bottom surface) $r=0.001$.
}\label{fig:EN_numinus:theta0_rs:vs:rho:alpha}
\end{figure}

In \Fig{fig:numinus:theta0:alphas} we also show the case of more realistic propagation loss:
(upper-right) $1\%$ loss ($\alpha=0.99$), and (lower-left) $5\%$ loss ($\alpha=0.95$). These plots indicate that although $\tnu_-(\rho)$ cannot be reduced identically to zero when realistic loss is present, it can still be substantially reduced below the value of a half on resonance $\theta=0$.
The lower-right plot in \Fig{fig:numinus:theta0:alphas} collects the graphs of $\tnu^{(\theta=0)}_-(\rho)$ for fixed $r=0.01$ for $\alpha=(1.0, 0.99, 0.95)$, showing the effects of operationally realistic propagation loss ($1\%, 5\%$) over that of the idealized lossless case ($\alpha=1$).
In \Fig{fig:EN_numinus:theta0_rs:vs:rho:alpha} we plot the log negativity $E_N(\rho,\alpha)$ for on mrr resonance $\theta=0$ as a function of $0\le\rho\le 1.0$ and $0.95\le\alpha\le 1.0$
for $r=(0.1, 0.01, 0.001)$ (compare \Fig{fig:numinus:theta0:alphas}). Again, this plot indicates that at realistic values of internal propagation loss ($\alpha<1$), a high cavity $Q$ ($\rho$ nearer to unity) enhances entanglement.

\subsection{The effect of only the $G$-terms on the log negativity}
It is instructive to compare the above symplectic eigenvalues
and log negativity plots which include signal-idler photon loss,
with the corresponding plots  using only the $G$-terms $\mA_G, \mA'_G, \mB_G, \mC_G$ in \Eq{sigma:xaxa}-\Eq{sigma:xayb}
in the symplectic eigenvalues (\Eq{nupm}), which we will label as $\tnu_{\pm}^{(G)}$.
Note that for the (ideal) lossless case $\alpha=1$, the $H$-terms $\mA_H, \mA'_H, \mB_H, \mC_H$ in the symplectic eigenvalues are identically zero.
For the case of loss $\alpha<1$, these $H$-terms are responsible
for driving the full symplectic eigenvalues (using both $G$ and $H$ terms) in \Eq{nupm}, towards the value of $\half$ where the log negative has the value zero. By considering the symplectic eigenvalues $\tnu_{\pm}^{(G)}$  comprised of only the classical-like loss (for $\alpha<1$) $G$-terms,
we can infer their influence for arbitrary values of $\alpha$. To give this a name we will refer to it as
`no quantum noise signal-idler photon loss.'

\begin{figure}[h t]
\begin{tabular}{cc}
\includegraphics[width=6.0in,height=2.00in]{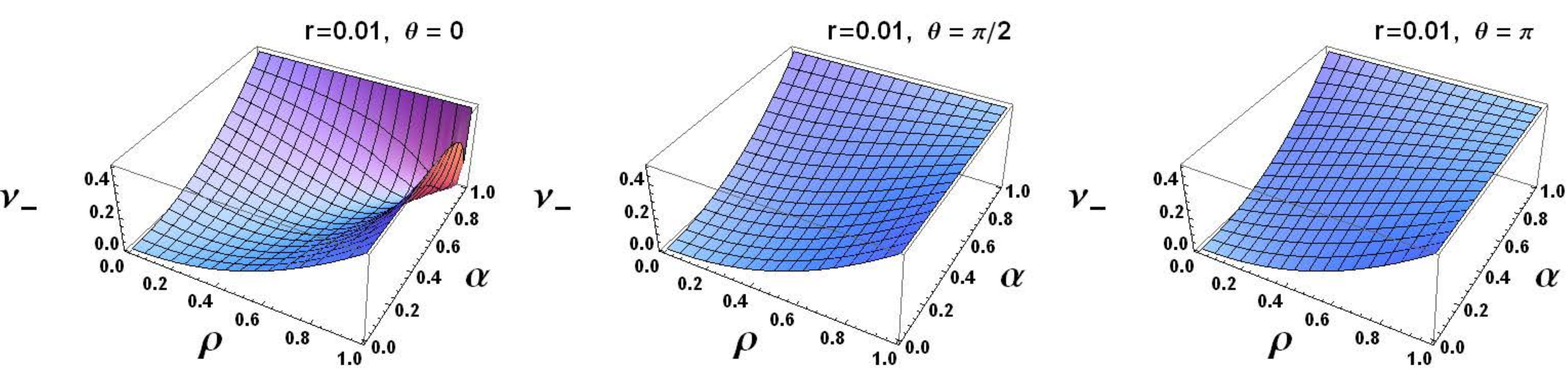} \\ 
\includegraphics[width=6.0in,height=2.00in]{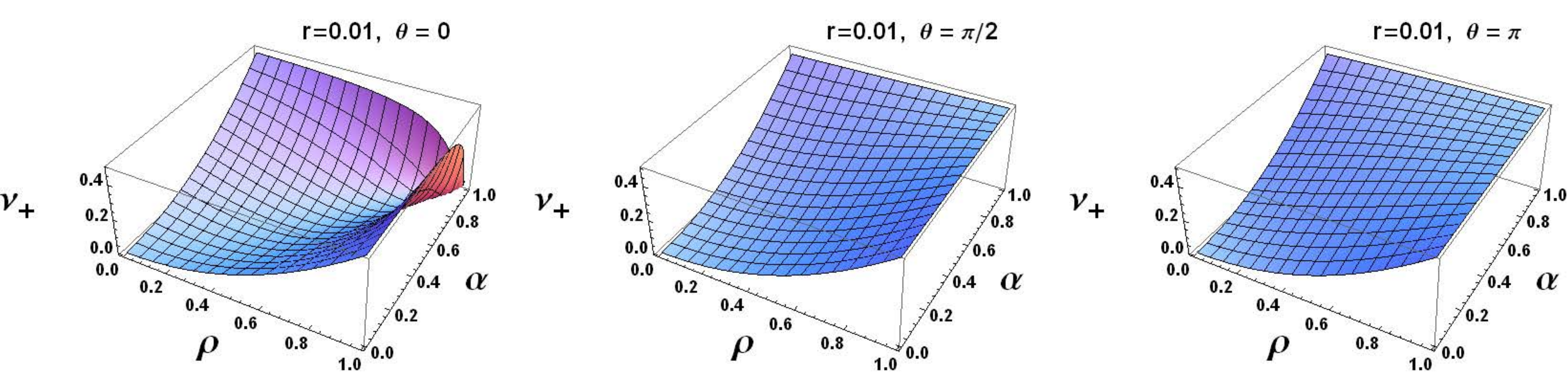} 
\end{tabular}
\caption{(top) $\tnu_-^{(G)}$ and (bottom) $\tnu_+^{(G)}$ (compare with \Fig{fig:nuplus:minus:thetas})
 for (left) on mrr resonance $\theta=0$,
(middle) slightly off mrr resonance $\theta=\pi/2$, (right) midway between mrr resonances $\theta=\pi$
for the case of no quantum noise signal-idler photon loss (i.e. $E_N$ computed without the $H$-terms in \Eq{aout:matrix:form}).
}\label{fig:nuplus:minus:thetas:onlyGterms}
\end{figure}

Using only the $G$-terms in \Eq{nupm} yields
\bea{nu:plus:minus:onlyGterms}
\tnu_{\pm}^{(G)} &\approx&
\half\,|S(\rho,\theta,\alpha)|^2\,f_0(\rho,\theta,\alpha)
\; \pm \;  r\,\alpha\,(1-\rho^2)\,|S(\rho,\theta,\alpha)|^4\,f_1(\rho,\theta,\alpha) \no
&+& \half\,r^2\,\alpha\,(1-\rho^2)\,|S(\rho,\theta,\alpha)|^4\,f_1(\rho,\theta,\alpha)
\eea
to be compared with \Eq{nu:plus:minus}. As in \Eq{nu:plus:minus},
$|S(\rho,\theta,\alpha)|^2 = |1-\rho\,\alpha\,e^{i\theta}|^{-2} = (1+\rho^2\,\alpha^2 - 2\,\rho\,\alpha\,\cos\theta)^{-1}$ is the modulus squared of the round trip circulation factor.
Here $f_0(\rho,\theta,\alpha) = |\alpha\,e^{i\,\theta} - \rho|^2 = \alpha^2 + \rho^2- 2\,\rho\,\alpha\,\cos\theta$, and
$f_1(\rho,\theta,\alpha)$ and $f_2(\rho,\theta,\alpha)$ are other polynomials of $\rho, \alpha$ and trigonometric functions of $\theta$.
In \Fig{fig:nuplus:minus:thetas:onlyGterms} we plot the full expressions for $\tnu_{\pm}^{(G)}$ for which \Eq{nu:plus:minus:onlyGterms} is numerically a very good approximation for $r<0.01$.
%
\begin{figure}[h t]
\includegraphics[width=6.5in,height=2.15in]{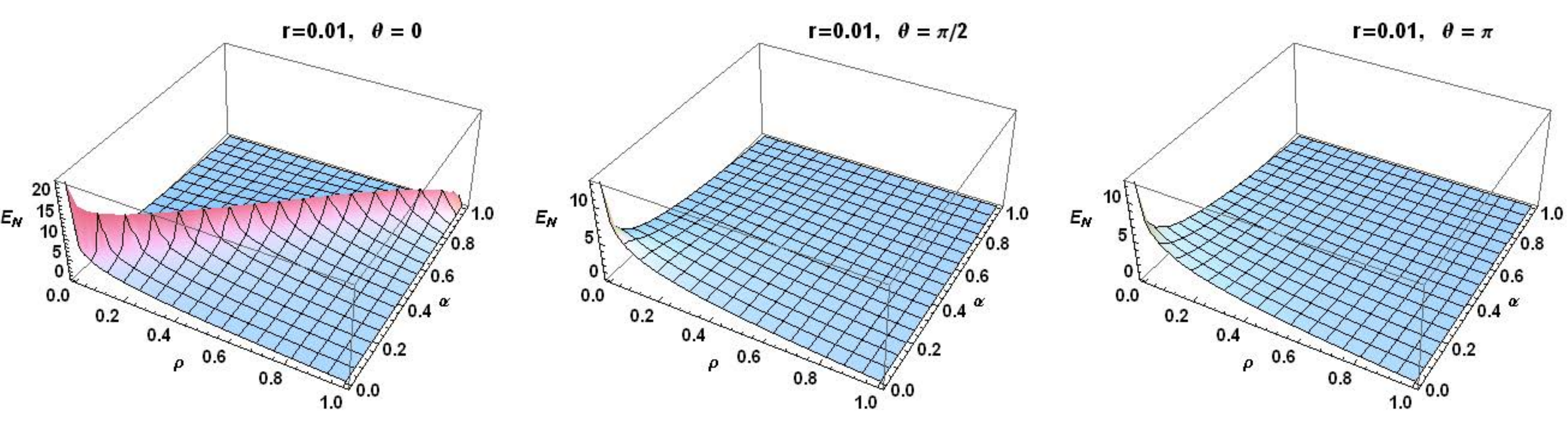} 
\caption{$E_N^{(out)}(\tnu_{\pm}^{(G)})$ for (left) on mrr resonance $\theta=0$, (middle) slightly off mrr resonance $\theta=\pi/2$,
(right) midway between mrr resonances $\theta=\pi$, for the case of
no quantum noise signal-idler photon loss
(i.e. $E_N$ computed without the $H$-terms in \Eq{aout:matrix:form}).
}\label{fig:EN:thetas:onlyGterms}
\end{figure}
%
\begin{figure}[h t]
\includegraphics[width=3.0in,height=3.5in]{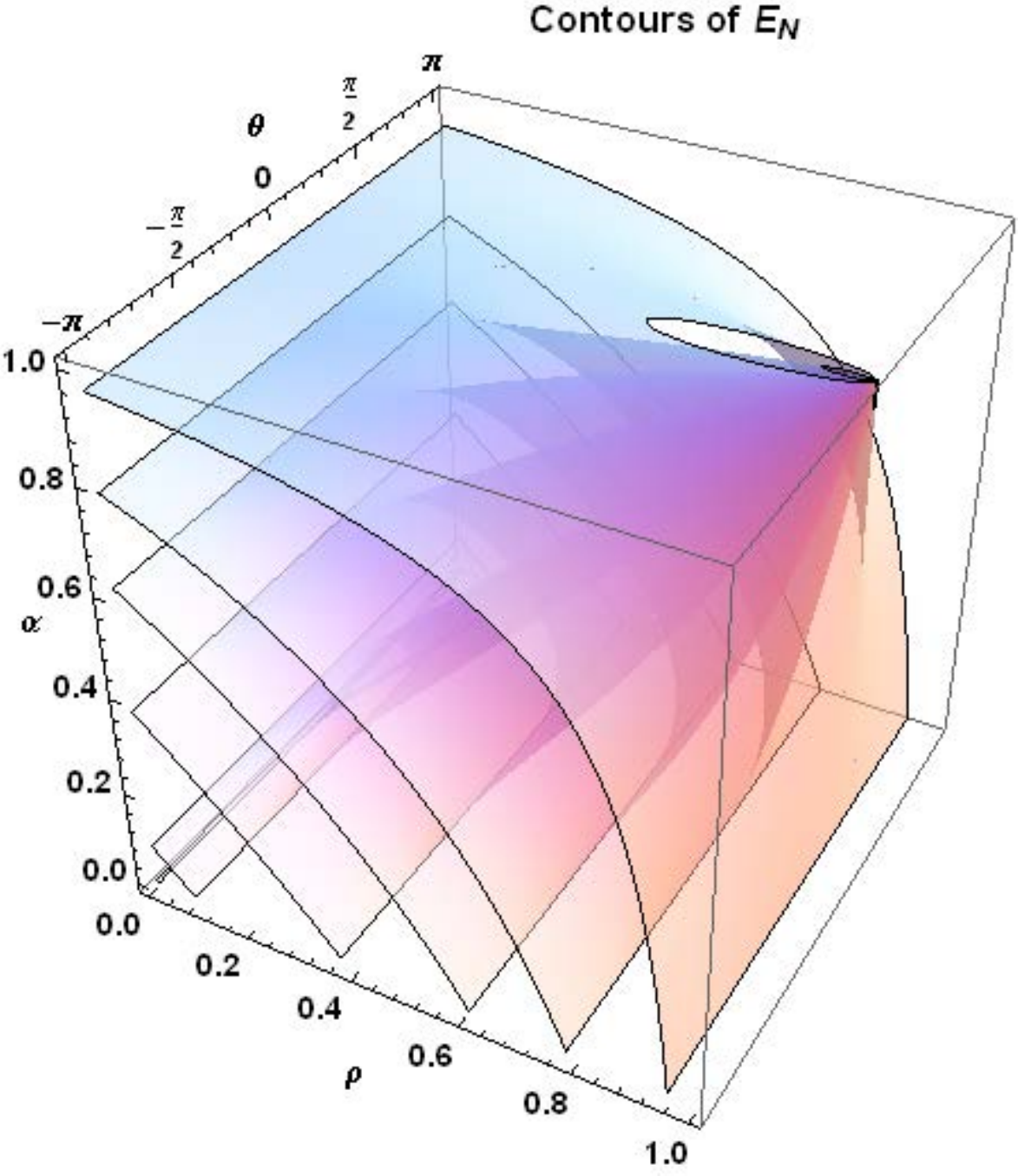} 
\caption{Contour plots (surfaces right to left in cube) of
(right) $E_N^{(out)}(\tnu_{\pm}^{(G)}) \in\{0.1, 0.5, 1.0, 2.0, 5.0, 10.0\}$
(corresponding to symplectic eigenvalues
$\tnu_{-}~\in~\{0.450, 0.300, 0.180, 0.068, 3.4\times 10^{-3}, 2.3\times 10^{-5}\}$)
for the case of no quantum noise signal-idler photon loss (i.e. $E_N$ computed without the $H$-terms in \Eq{aout:matrix:form}).
}\label{fig:contours:numinus_EN:onlyGterms}
\end{figure}
%
As opposed to the full symplectic eigenvalues $\tnu_{\pm}$ in \Eq{nu:plus:minus}, we see that both $\tnu_{\pm}^{(G)} < \half$ over the entire range of parameters, indicating that a degree of entanglement is always present via the log negativity \Eq{logNeg:nupm} as indicated in
\Fig{fig:EN:thetas:onlyGterms} and \Fig{fig:contours:numinus_EN:onlyGterms}.
Using the full expression for the $G$-term symplectic eigenvalues we have
$\tnu_{\pm}^{(G)}\rightarrow 0$ in the limit $\rho,\alpha\rightarrow 0$,
which is why  the contour surfaces of high log negativity in \Fig{fig:contours:numinus_EN:onlyGterms} converge to the lower left edge of the cube. This is also evident in \Fig{fig:EN:thetas:onlyGterms} where the log negativity peaks for
$\rho,\alpha\rightarrow 0$,
where correspondingly in \Fig{fig:nuplus:minus:thetas:onlyGterms} the symplectic eigenvalues $\tnu_{\pm}^{(G)}$ are zero.
%
The point is that  the symplectic eigenvalues $\tnu_{\pm}^{(G)}$ using only $G$ terms
favor large entanglement for parameter values
$\rho,\alpha\rightarrow 0$, while the physical case of signal-idler photon loss (including both $G$ and $H$ terms
in the symplectic eigenvalues) $\tnu_{\pm}$ favors the parameter regime of low internal propagation loss and high cavity Q, $\rho,\alpha\rightarrow 1$.
The $G$ terms in the full symplectic eigenvalues drive $\tnu_\pm$ towards values less than a half, even towards zero, while the $H$ terms drive  $\tnu_\pm$ towards values near a half. Depending on the value of dimensionless pump parameter $r=g\,\alpha_p\,T$ and the mrr parameters $\rho,\theta,\alpha$, a balance can be reached between the competing $G$ and $H$ terms such that a degree of entanglement is preserved by the mrr, even in the case of signal-idler photon loss.

\section{Summary and Discussion}\label{sec:summary:discussion}
In this work we have investigated the entanglement of the output signal-idler squeezed vacuum state in the Heisenberg picture as a function of the coupling and internal propagation loss parameters of a microring resonator.
We constructed the operator expression that produces the output squeezed vacuum state in the presence of noise.
We also constructed the unitary operator that evolves the input modes to the output modes
$\vec{a}_{out}(\om) = G(\om) \, \vec{a}_{in}(\om) + H(\om)\, \vec{f}(\om)$ (see \Eq{aout:matrix:form}) with the inclusion of loss. Since the mrr is essentially a linear optical beam splitter with passive feedback, Gaussian input states are evolved to Gaussian output states even in the presence of noise. Hence, we investigated the entanglement of the mrr output two-mode squeezed state using the log negativity and the symplectic structure of mixed Gaussian states. We showed that the transfer matrix $G(\om)$ which encodes the classical phenomenological loss (for $\alpha<1$) pulls the symplectic eigenvalues of the covariance matrix of the mixed Gaussian state towards zero, where the log negativity is large, indicating strong entanglement. On the other hand, the noise matrix $H(\om)$ pulls the eigenvalues towards the value of $1/2$, where the log negativity, and hence the entanglement is small. We investigated the role of the (constant) driving pump and nonlinear coupling constant $g\alpha_p$ on the entanglement of the output mixed Gaussian squeezed state, and showed that depending on its strength, the symplectic eigenvalues can be driven towards zero for certain values of the self-coupling (`reflection') parameter $\rho$ when propagation losses are small ($\alpha$ near unity).

This work represents the most recent step toward our overarching goal of developing a comprehensive theoretical framework and computational tool kit for the design and optimization of a class of scalable, on-chip linear quantum optical information processing devices. Previously, we have (i) examined the quantum dynamics of a single bus microring resonator \cite{Hach:2010}, (ii) proposed and analyzed a 'fundamental circuit' element for this class of devices \cite{Hach:2014}, and (iii) extended the analysis of the fundamental circuit element to examine its response in the presence of quantum noise \cite{Alsing_Hach:2016}. Specifically, in references \cite{Hach:2010,Hach:2014,Alsing_Hach:2016} we demonstrate theoretically advantageous enhancements of the operating parameter spaces of the devices we consider owing to the Passive Quantum Optical Feedback (PQOF) that is a signature feature of the architecture for this class of device. In this paper, and in the first paper in this two paper sequence (AH-I), we have extended the analysis to include on-chip, intra-ring photon generation via the processes of SPDC and SFWM. In this paper specifically, we have analyzed the competitive effects due to (i) (amplitude) attenuation noise and (ii) quantum noise arising from coupling with environment on the level of entanglement present in states transmitted a single bus device featuring PQOF. These results are instrumental to understanding the practical quantum information processing capabilities devices of this sort under more realistic operating conditions.

Our current and future work is focused upon using the theoretical and computational tools we have developed so far in \cite{Hach:2010,Hach:2014,Alsing_Hach:2016,Alsing_Hach:2017a} and this current work to inform the design and to optimize the function of devices of high impact for Linear Quantum Optical Information processing, such as the Knill-Laflamme-Milburn (KLM) CNOT gate \cite{KLM:2001}. Further, we are investigating larger networks of directionally coupled silicon nanophonotonic waveguide/mrr arrays for possible quantum advantages with respect to communications, sensing and metrology \cite{Preble:2015,Vernon:2017}.

\begin{acknowledgments}
PMA, would like to acknowledge support of this work from
Office of the Secretary of Defense (OSD) ARAP QSEP program, and thank
J. Schneeloch and M. Fanto for helpful discussions.
EEH would like to acknowledge support for this work was provided by the Air Force Research
Laboratory (AFRL) Visiting Faculty Research Program
(VFRP) SUNY-IT Grant No. FA8750-13-2-0115.
Any opinions, findings and conclusions or recommendations
expressed in this material are those of the author(s) and do not
necessarily reflect the views of Air Force Research Laboratory.
\end{acknowledgments}

\bibliography{rr_losses_refs}
\end{document}